\documentclass[lettersize,doublecolumn]{IEEEtran}
\IEEEoverridecommandlockouts
\usepackage{lineno}
\usepackage{hyperref}
\usepackage{cite}
\usepackage{amsmath,amssymb,amsfonts,cases} 
\usepackage{amsmath}
\usepackage{amsthm}
\usepackage{subfig}

\usepackage{graphicx}
\usepackage{textcomp}
\usepackage{xcolor}
\usepackage{graphicx}
\usepackage{float}
\usepackage{amsmath,mleftright}
\usepackage{amsfonts,amssymb}
\usepackage{mathrsfs}
\usepackage{mathtools}
\usepackage[linesnumbered,ruled,vlined]{algorithm2e}
\usepackage{bm}
\usepackage{multirow}
\usepackage{array}
\usepackage{amssymb}
\usepackage{amsmath}
\usepackage{cite}
\usepackage{url}
\usepackage{xcolor}
\usepackage{cite,graphicx,amsmath,amssymb}
\usepackage{fancyhdr}
\usepackage{mdwmath}
\usepackage{mdwtab}
\usepackage{caption}
\usepackage{amsthm}
\usepackage{setspace}
\usepackage{bm}
\usepackage{mathtools}
\usepackage{dsfont}
\usepackage{bbm}
\usepackage{framed}
\allowdisplaybreaks[4]
\newtheorem{remark}{Remark}
\newtheorem{theorem}{Theorem}

\newtheorem{lemma}{Lemma}

\newtheorem{corollary}{Corollary}

\makeatletter
\newcommand{\biggg}{\bBigg@{3}}
\newcommand{\Biggg}{\bBigg@{3.5}}
\makeatother
\makeatletter
\renewcommand{\maketag@@@}[1]{\hbox{\m@th\normalsize\normalfont#1}}%
\makeatother
\def\BibTeX{{\rm B\kern-.05em{\sc i\kern-.025em b}\kern-.08em
    T\kern-.1667em\lower.7ex\hbox{E}\kern-.125emX}}
    \expandafter\def\expandafter\normalsize\expandafter{%
    \normalsize%
    \setlength\abovedisplayskip{4pt}%
    \setlength\belowdisplayskip{4pt}%
    \setlength\abovedisplayshortskip{2pt}%
    \setlength\belowdisplayshortskip{2pt}%
}
\begin{document}
\title{\vspace{-0.5em}Pinching-Antenna Assisted Sensing: \\A Bayesian Cram\'{e}r-Rao Bound Perspective}

\author{Hao Jiang, Chongjun Ouyang, Zhaolin Wang, Yuanwei Liu,~\IEEEmembership{Fellow,~IEEE,} \\
Arumugam Nallanathan,~\IEEEmembership{Fellow,~IEEE,} and Zhiguo Ding,~\IEEEmembership{Fellow,~IEEE} \vspace{-1em}
\thanks{
H. Jiang, C. Ouyang, and A. Nallanathan are with the School of Electronic Engineering and Computer Science, Queen Mary University of London, London E1 4NS, U.K. (e-mail: \{hao.jiang, c.ouyang, a.nallanathan\}@qmul.ac.uk).
\\
Z. Wang and Y. Liu are with the Department of Electrical and Electronic Engineering, The University of Hong Kong, Hong Kong (e-mail: \{zhaolin.wang, yuanwei\}@hku.hk).
\\
Z. Ding is with Khalifa University, Abu Dhabi, UAE, and the University of Manchester, Manchester, M1 9BB, U.K. (e-mail: zhiguo.ding@manchester.ac.uk).
}\vspace{-1em}
}
\maketitle
\begin{abstract}
The fundamental sensing limit of pinching-antenna systems (PASS) is studied from a Bayesian Cramér-Rao bound (BCRB) perspective.
Compared to conventional CRB, BCRB is independent of the exact values of sensing parameters and is not restricted by the unbiasedness of the estimator, thus offering a practical and comprehensive lower bound for evaluating sensing performance.
A system where multiple targets transmit uplink pilots to a single-waveguide PASS under a time-division multiple access (TDMA) scheme is analyzed.
For the single-target scenario, our analysis reveals a unique mismatch between the sensing centroid (i.e., the optimal PA position) and the distribution centroid (i.e., the center of the target's prior distribution), underscoring the necessity of dynamic PA repositioning.
For the multi-target scenario, two target scheduling protocols are proposed: 1) pinch switching (PS), which performs separate pinching beamforming for each time slot, and 2) pinch multiplexing (PM), which applies a single beamforming configuration across all slots.
Based on these protocols, both the total power minimization problem under a BCRB threshold and the min-max BCRB problem under a total power constraint are formulated.
By leveraging Karush-Kuhn-Tucker (KKT) conditions, these problems are equivalently converted into a search over PA positions and solved using an element-wise algorithm.
Numerical results show that i)~PASS, endowed with large-scale reconfigurability, can significantly enhance the sensing performance compared with conventional fixed-position arrays, and ii)~PS provides more robust performances than PM at the cost of higher computational complexity.
\end{abstract}

\section{Introduction}
Beneath the everlasting quest for the advancement of wireless technologies, multiple-input multiple-output (MIMO) technology plays an indispensable role by offering high array gain, spatial multiplexing gain, and diversity gain, thus heralding a future of high-speed connectivity \cite{lu2014mimo}.
Nowadays, MIMO has evolved into a more flexible form, namely the \emph{reconfigurable antenna}, characterized by its ability to proactively manipulate wireless channels \cite{ouyang2025capacity}.

Added as the outer layer of the conventional digital or hybrid beamforming configurations, reconfigurable antennas enable performance enhancement by judiciously altering the EM properties of wireless signals, while reducing hardware cost and energy consumption, a scheme termed as \emph{tri-hybrid beamforming} \cite{castellanos2025embracing}.
For instance, reconfigurable intelligent surfaces (RISs) introduce virtual line-of-sight (LoS) links between transceivers \cite{huang2019reconfigurable}, while movable antennas (MA) and fluid antennas (FA) exploit antenna repositioning to reconfigure wireless channels flexibly \cite{zhu2025tutorial, new2025tutorial}.
Instead of passively adapting to propagation conditions, reconfigurable antennas substantially enhance wireless transmissions across a wide performance spectrum, including array gain and interference cancellation, diversity gain, and multiplexing gain \cite{new2024fluid, wong2022fluid}.
Furthermore, since communication functionality has largely been fulfilled in practice, the sensing functionality deserves investigation, as it represents a key conceptual tool for intelligent connectivity \cite{liu2022survey}.
To demonstrate the sensing potential of reconfigurable antennas, \cite{ma2024movable} studied the uplink angle estimation problem in the context of MA and revealed the dependence of sensing accuracy on array geometry, indicating that dedicated antenna repositioning can improve sensing performance.
This work also shows that MA can mitigate angle estimation ambiguity compared with conventional fixed-position arrays.

However, the conventional reconfigurable antenna technologies can only offer limited flexibility, typically on the scale of a few wavelengths \cite{liu2025pinching, ouyang2025array}.
As such, existing reconfigurable antennas are suitable for mitigating small-scale fading rather than more dominant large-scale fading.
To overcome this limitation, a pinching-antenna system (PASS) was proposed by NTT DOCOMO in 2021 \cite{fukuda2022pinching, ding2025flexible}.
PASS consists of long, low-attenuation waveguides and several signal-leaking particles, referred to as pinching antennas (PAs).
In contrast to conventional antenna apertures, typically limited to tens or hundreds of wavelengths, the waveguide can extend to tens of meters, allowing low-attenuation signal propagation inside while preventing energy dissipation into free space \cite{xu2025pinching}.
By attaching PAs to the waveguide and judiciously adjusting their positions, in-waveguide signals can be radiated at desirable locations, thereby enabling large-scale reconfigurability \cite{ding2025flexible}.
As such, wireless communications are converted from ``last mile" to ``last meter", analogous to ``PASS" signals to users over short distances \cite{liu2025pinching}. 

Endowed with higher flexibility, PASS can significantly reconfigure wireless channels and facilitate wireless transmissions \cite{ouyang2025capacity, wang2025modeling}, thus fostering a wide range of application scenarios, as detailed in a recent tutorial paper \cite{liu2025pinching}. 
Despite the validated communication enhancement, PASS can also benefit sensing functionality in the following ways:
i)~\emph{High-Resolution Sensing}: thanks to the long-extending waveguides, the large aperture size created by PASS improves sensing resolution \cite{rusek2013scaling}, enabling the precise distinction of closely spaced targets; ii)~\emph{Full-Dimensional Sensing}: accompanied by larger aperture sizes, near-field effects become dominant, enabling polar-domain localization \cite{jiang2025nearfield}; and iii)~\emph{Low-Cost Sensing}: compared to fixed-position array (FPA) counterparts, PASS-based sensing can reduce the number of power-hungry RF chains, thereby lowering implementation cost.
These benefits of PASS-based sensing have been explored in some initial attempts.
In particular, the authors of \cite{qin2025joint} and \cite{ zhang2025integrated} first introduced PASS into integrated sensing and communication (ISAC) systems, where the received signal-to-noise ratio (SNR) at targets was utilized as the sensing metric.
Although their results demonstrated the superiority of PASS for sensing, this sensing metric, i.e., received SNR at targets, only reflects illumination power and may not be directly related to actual estimation performance.
To address this issue, the Cramér-Rao bound (CRB) is exploited instead, as it provides the lower bound on the estimation error variance for any unbiased estimator \cite{kay1993estimation, liu2022cramer}.
As such, minimizing the CRB can reduce the variance of estimation error, thereby improving sensing accuracy.
Building on this, the authors of \cite{ding2025pinchingisac} derived the CRB for uplink PASS-based sensing scenarios, while minimizing the CRB was addressed in \cite{wang2025wireless} using a particle swarm optimization (PSO)-based algorithm.  
It is important to note that the results in \cite{ding2025pinchingisac} demonstrated an inherent conflict between the communication and sensing functionalities of PASS: directly placing PAs above the users enhances throughput, but simultaneously degrades sensing accuracy.
In parallel, the authors of \cite{bozanis2025cramer}, \cite{li2025pinching}, and \cite{khalili2025pinching} considered a round-trip sensing channel and utilized uniform linear arrays (ULAs) for echo signal reception.
Moreover, the authors of \cite{ouyang2025rate} analyzed the information-theoretic rate region of PASS-aided ISAC systems, revealing a fundamental tradeoff between communication rate and sensing rate.
    
    Although the former works have laid a solid foundation for PASS-aided sensing, existing research still faces challenges in the following aspects.
    First, current studies on PASS sensing are primarily based on the CRB, a function parameterized by the ground-truth parameters of the sensing target, such as position \cite{kay1993estimation, xu2024mimo, hou2024optimal}.
    As such, this bound is more suitable for evaluating sensing performance than for serving as an optimization objective, since these parameters are unavailable at the estimation stage.
    Second, the majority of existing work focuses on CRB minimization using heuristic methods widely applied to conventional reconfigurable antennas, such as PSO \cite{wang2025wireless}, thereby overlooking the fundamental differences introduced by PASS, including the dual phase shifts that occur both inside and outside the waveguides.
    This oversight prevents current work from providing insightful system design guidelines for PASS, such as identifying the optimal PA positions in a sensing setup.
    Last but not least, due to the limited attention to the multi-target scenario, a dedicated protocol design is needed to enable effective multi-target sensing with a single waveguide and a single RF chain. 

    To address the above challenges, we utilize the Bayesian CRB (BCRB) to relax the conventional CRB's reliance on the exact values of the sensing parameters.
    This is because the BCRB leverages prior information on the targets' distributions \cite{xu2024mimo, hou2024optimal, dauwels2005computing}.
    Such information can be obtained from historical statistical data.
    Subsequently, we derive the optimal PA position for the one-dimensional position uncertain case, which reveals the mismatch between sensing and communication functionalities in terms of PA placement.
    Furthermore, we design a pair of protocols for PASS-based sensing, namely \emph{pinch switching} and \emph{pinch multiplexing}, to provide a scalable solution for the multi-target sensing scenario.
    The main contributions are summarized as follows:
    \begin{itemize}
        \item We consider a PASS-aided uplink multi-target sensing scenario, where targets transmit uplink pilots to a PASS receiver with a single waveguide in a time-division multiple access (TDMA) manner.
        We derive the BCRB as the sensing performance metric, which is independent of the exact sensing parameters. 
        
        \item We first consider a simplified setup with a single PA and a single target, where only one-dimensional position uncertainty is analyzed to gain insights.
        Building on this, we exploit an approximation method to identify the optimal PA position, and its tightness is further examined. 
        The optimal PA position reveals a mismatch between the distribution centroid and the sensing-sensitive centroid.
        This unique feature of PASS-based sensing underscores the necessity of judicious PA position design. 
        Then, an element-wise algorithm is proposed for the general multi-pinch scenario.
        
        \item We then extend to the multi-target case.
        To enable multi-target sensing, a pair of sensing protocols is proposed: pinch switching (PS) and pinch multiplexing (PM).
        The former performs dedicated pinching beamforming for each target on its assigned time slot, whereas the latter performs pinching beamforming once for all targets collectively.
        Under these protocols, we formulate the total power minimization problem under a targeted BCRB constraint and the min-max BCRB problem under a total power constraint.
        To solve these problems, we utilize the Karush–Kuhn–Tucker (KKT) conditions to transform them into equivalent forms, where only the PA positions need to be optimized.
        Then, we develop an element-wise algorithm to effectively perform PA position optimization.
        For the high-SNR regime, we further present an alternative algorithm for the min-max BCRB problem to reduce computational complexity.
        
        \item We present numerical results to validate our derivations and the effectiveness of PASS-based sensing.
        The simulation results demonstrate that: i)~the proposed algorithms can effectively solve the formulated problems when compared with heuristic solutions, ii)~with enhanced flexibility, PASS can offer significant performance gains compared to conventional FPA, and iii)~due to the increased optimization dimensions, i.e., treating each time slot separately, PS consistently outperforms its PM counterpart and also shows better robustness, albeit at the cost of higher computational complexity. 
    \end{itemize}
    
    The rest of the paper is organized as follows:
    Section \ref{sect:system_model} presents the PASS-based sensing system model and elaborates on the derivations of the BCRB expressions. 
    Section \ref{sect:single_target} considers a simplified single-target scenario to gain insights into this problem, where the unique mismatch is revealed. 
    Then, Section \ref{sect:multi_target} focuses on more challenging multi-target scenarios and proposes the algorithm designs for both the total power minimization and min-max BCRB problems under PS and PM protocols, respectively.
    Numerical results are provided in Section \ref{sect:simulation_results}, and conclusions are drawn in Section \ref{sect:conclusion}.

    \textit{Notations:}
    Scalars, vectors, and matrices are denoted by the lower-case, bold-face lower-case, and bold-face upper-case letters, respectively.
    $\mathbb{C}^{M \times N}$ and $\mathbb{R}^{M \times N}$ denote the space of $M \times N$ complex and real matrices, respectively.
    $(\cdot)^\mathrm{T}$, $(\cdot)^*$, and $(\cdot)^\mathrm{H}$ denote the transpose, conjugate, and conjugate transpose, respectively.
    $|\cdot|$ represents absolute value.
    $\mathbb{E}\{\cdot\}$ denotes the expectation operation over a random variable $(\cdot)$, $\mathrm{tr}\{\cdot\}$ denotes the trace of matrix $(\cdot)$, and $\Re \{\cdot\}$ denotes the operation to extract the real part of $(\cdot)$.
    For a vector $\mathbf{a}$, $[\mathbf{a}]_i$ and $\left\| \mathbf{a} \right\|_2$ denote the $i$-th element and $2$-norm, respectively.
    $\mathrm{j}=\sqrt{-1}$ denotes the imaginary unit.
    
\section{System Model}\label{sect:system_model}
\begin{figure}[t!]
\centering
\includegraphics[width=0.4\textwidth]{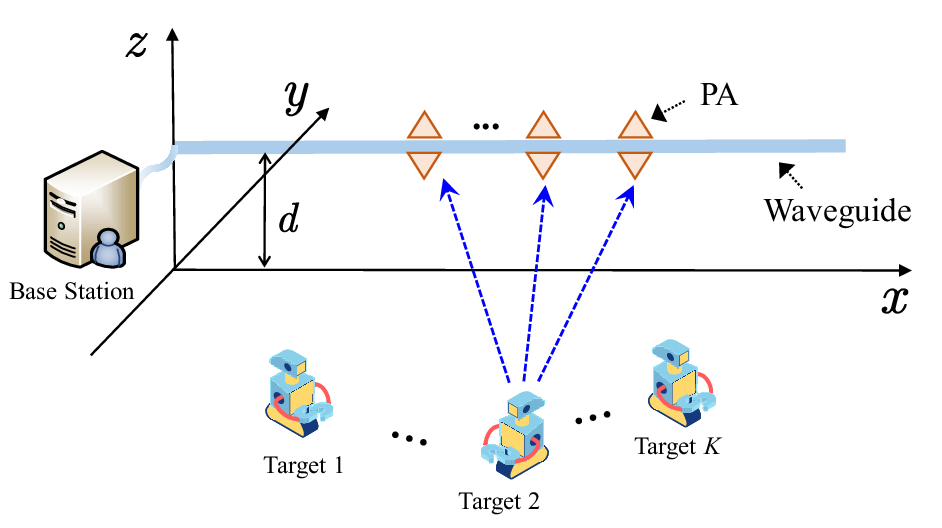}
\caption{Illustration of a wireless sensing system with pinching antennas.}
\label{fig:syst}
\end{figure}
In this work, an uplink sensing framework is investigated, where the $K$ targets transmit uplink pilot signals to a PASS receiver, i.e., base station (BS), equipped with a waveguide and $M$ PAs.
Without loss of generality, it is assumed that the waveguide is placed along the $x$-axis with the feed point located at $x_0=0$.
Additionally, the waveguide is parallel to the $XOY$ plane at a height of $d$.
The feasible PA positions are confined within $\mathcal{S}=\{x: 0 \le x\le x_{\max}\}$, where $x_{\max}>0$ denotes the waveguide length.
A common narrow band is used for all the uplink transmissions. 
Since PASS is promising in high-frequency bands that are dominated by LoS, we adopt LoS channels for analytical tractability \cite{wang2025wireless, li2025pinching, ding2025pinching}.

\subsection{Signal Model and Sensing Metric}
Let the position of the $m$-th PA be specified as $\mathbf{p}_{m}=[x_m, 0, d]^{\rm T} \in \mathbb{R}^{3 \times 1}$.
As shown in Fig. \ref{fig:syst}, there are $K$ sensing targets located on the $XOY$ plane, where the position of the $k$-th sensing target is specified by $\mathbf{r}_k=[r_x, r_y, 0]^{\rm T} \in \mathbb{R}^{3 \times 1}$ with $k\in\mathcal{K}\triangleq\{1,2,...,K\}$.
Letting the uplink pilot of the $k$-th sensing target be $s_k$ with $\mathbb{E}\{|s_k|^2\}=P_k$, the receive signal at the BS is given by
\begin{align}
    y&=\sum\nolimits_{k=1}^K{\mathbf{g}^{\mathrm{T}}(\mathbf{x};\mathbf{r}_k)\mathbf{h}_k(\mathbf{x};\mathbf{r}_k)s_k}+n\notag\\
    &=\sum\nolimits_{k=1}^K{\mathbf{f}_k(\mathbf{x};\mathbf{r}_k)s_k}+n, \label{eq:uplink_signal}
\end{align}
where $\mathbf{x} \triangleq [x_1, x_2, ..., x_M]^{\mathrm{T}} \in \mathbb{R}^{M \times 1}$ denotes the $x$-coordinate vector for the PAs, $n \sim \mathcal{CN}(0, M\tilde{\sigma}^2)$ denotes the additive Gaussian noise with zero mean and variance of $M\tilde{\sigma}^2$, $\mathbf{g}(\mathbf{x}) \in \mathbb{C}^{M \times 1}$ is the in-waveguide channel vector, and $\mathbf{h}_k(\mathbf{x}) \in \mathbb{C}^{M \times 1}$ represents the free-space channel vector.
For simplicity, we denote the overall noise power as $\sigma^2 \triangleq M \tilde{\sigma}^2$.
More specifically, the free-space channel vector from the $k$-th target to the PAs can be written as 
\begin{align}
    \mathbf{h}_k(\mathbf{x};\mathbf{r}_k) =\left[ \frac{\eta e^{-\mathrm{j}k_0r_{k,1}}}{r_{k,1}},...,\frac{\eta e^{-\mathrm{j}k_0r_{k,M}}}{r_{k,M}} \right] ^{\mathrm{T}}\in \mathbb{C} ^{M\times 1}, \label{eq:free_channel_vector}
\end{align}
where $\eta = \lambda/(4 \pi)$ is the path-loss factor with $\lambda$ being the carrier wavelength,  $k_0=2\pi/\lambda$ denotes the wavenumber in the free space, and $r_{k,m}=\left\| \mathbf{r}_k-\mathbf{p}_m \right\| _2$ is the Euclidean distance between the $k$-th target and the $m$-th PA.
After being received at the PAs, the signals will be guided through the waveguide, leading to in-waveguide phase shifts.
This effect is modeled by the in-waveguide channel vector, which is given by
\begin{align}
    \mathbf{g}(\mathbf{x};\mathbf{r}_k) =\left[ e^{-\mathrm{j}k_{\mathrm{w}}x_1},..., e^{-\mathrm{j}k_{\mathrm{w}}x_M} \right] ^{\mathrm{T}}\in \mathbb{C} ^{M\times 1}, \label{eq:in_waveguide_channel_vector}
\end{align}
where $k_{\mathrm{w}}=2\pi/\lambda_{\rm w}$ denotes the in-waveguide wavenumber with $\lambda_{\rm w} \triangleq \lambda / n_{\rm eff}$ being the in-waveguide wavelength, and $n_{\rm eff}=1.4$ is the effective refractive index of the waveguide.
To circumvent the interference between the uplink pilots sent by targets, we adopt TDMA when the sensing targets transmit uplink pilots to the BS.
More specifically, each target is allocated to a dedicated time slot for uplink pilot transmission.
This work considers an indoor scenario. 
In this case, the GPS data is inapplicable for positioning purposes.
Therefore, the positions of the targets are extracted from the uplink pilots they send, similar to \cite{wang2025wireless}.

To evaluate the sensing performance, the mean square error (MSE) is typically utilized.
Since all targets are located on the $XOY$ plane, their vertical positions are fixed at zero and therefore do not need to be estimated.
For the $k$-th sensing target, the MSE between the ground-truth position $\mathbf{r}_k=[{x}_k, {y}_k]^{\mathrm{T}}\in\mathbb{R}^{2 \times 1}$ and the estimated position $\hat{\mathbf{r}}_{k}=[\hat{x}_k, \hat{y}_k]^{\mathrm{T}}\in\mathbb{R}^{2 \times 1}$ is given by
\begin{align}
    \mathrm{MSE}_k\left( \mathbf{r}_k,\hat{\mathbf{r}}_k \right) &=\mathbb{E} _{\mathbf{r}_k}\left\{ \left\| \mathbf{r}_k-\hat{\mathbf{r}}_k \right\| _{2}^{2} \right\} \notag \\
    &=\mathbb{E} _{\mathbf{r}_k}\left\{ \left( x_k-\hat{x}_k \right) ^2+\left( y_k-\hat{y}_k \right) ^2 \right\}, \label{eq:mse}
\end{align}
where $\mathbf{r}_k \in \mathcal{R} _{\mathbf{r}_k}$ and $\mathcal{R} _{\mathbf{r}_k}$ denotes the position distributions of the $k$-th sensing target.
However, the MSE expression in \eqref{eq:mse} is mathematically intractable and can only be evaluated via empirical trials.
To this end, we resort to the Bayesian CRB denoted by $\text{BCRB}_k(\mathbf{x})$ for $\forall~k$, which serves as a fundamental lower bound of MSE, i.e., 
\begin{align}
    \mathrm{MSE}_k \ge  \mathrm{BCRB}_k(\mathbf{x}).
\end{align}
Before proceeding with the derivation of the BCRB, we first explain the motivation for adopting the BCRB, highlighting its difference from the conventional CRB.
Compared to conventional CRB \cite{liu2022cramer, jiang2024cramer, li2008range}, there are two important motivations for us to adopt the BCRB:
1)~\emph{Unknown Parameter Free}: The conventional CRB is parameterized by the unknown parameters, making it more suitable for target tracking and presence detection of the sensing target \cite{liu2022cramer}.
On the contrary, BCRB utilizes the prior distributions of the unknown parameters rather than their precise values, making it a more favorable choice for practical usage; and 2)~\emph{General Bound for Different Estimators}: the conventional CRB is the lower bound for unbiased estimators \cite{kay1998fundamentals}.
In contrast, the BCRB is a global CRB, which can be used to evaluate the theoretical performance limit for any estimator \cite{van2004detection}.

Given the above advantages, we will proceed with the derivation of BCRB.
Similar to the conventional CRB, BCRB is also defined as the inverse of the Bayesian Fisher information matrix (BFIM) \cite{scope2025bayesian}.
Therefore, for the $k$-th sensing target, its BCRB is specified by 
\begin{align}
    \mathrm{BCRB}_k\left( \mathbf{x} \right) =\mathrm{tr}\left\{\mathbf{J}_{\mathbf{r}_k}^{-1}\left( \mathbf{x} \right) \right\}, \label{eq:fim}
\end{align}
where $\mathbf{r}_k\triangleq[x_k, y_k]^{\mathrm{T}} \in \mathbb{R}^{2 \times 1}$ is the unknown parameter vector for the $k$-th sensing target,
In \eqref{eq:fim}, the BFIM for the $k$-th sensing target, i.e., $\mathbf{J}_{\mathbf{r}_k} \in \mathbb{R}^{2 \times 2}$, can be expressed as \cite{shen2010fundamental}
\begin{align}
    \mathbf{J}_{\mathbf{r}_k}^{}\left( \mathbf{x} \right) =\underset{\mathrm{Observation}}{\underbrace{\check{\mathbf{J}}_{\mathbf{r}_k}^{}\left( \mathbf{x} \right) }}+\underset{\mathrm{Prior}}{\underbrace{\hat{\mathbf{J}}_{\mathbf{r}_k}^{} }},
\end{align}
where $\check{\mathbf{J}}_{\mathbf{r}_k}^{}\left( \mathbf{x} \right) $ denotes the FIM associate real observation captured by \eqref{eq:uplink_signal}, while $\hat{\mathbf{J}}_{\mathbf{r}_k}^{}$ denotes the prior knowledge on the distribution of the $k$-th target.
In what follows, we will elaborate on the derivations of the FIMs regarding the observation and the prior parts, respectively.

As a prerequisite step, the distributions of target locations need to be identified.
Here, we assume that the position of targets conforms to the Gaussian mixture model (GMM), which can be regarded as a weighted summation of multiple independent Gaussian distributions.
The reason why this particular distribution is chosen lies in the fact that location-scale mixtures of a continuous probability density function (PDF) can uniformly approximate any continuous PDF on a compact set \cite{nguyen2023approximation}.
Thus, when the positions of sensing targets conform to other distributions, one can approximate this distribution with GMM.
As the sensing targets access the BS in a TDMA manner, we can analyze each time slot independently, so that we drop the index of different sensing targets, i.e., $k$, for brevity.
In particular, for an arbitrary target, the PDFs for its $x$-coordinate and $y$-coordinate are given by
\begin{align}
    \begin{cases}
	p_x\left( r_x \right) =\sum\nolimits_{l_1=1}^{L_1}{\phi _{x, l_1} \mathcal{N} (r_x;u_{x,l_1},\sigma _{x,l_1}^{2})},\\
	p_y\left( r_y \right) =\sum\nolimits_{l_2=1}^{L_2}{\phi _{y, l_2}\mathcal{N} (r_y;u_{y,l_2},\sigma _{y,l_2}^{2})},\\
\end{cases}
\end{align}
where $\phi _{x/y,l_1/l_2} \in [0, 1] $ denotes the weight coefficients for the $l_1/l_2$-th Gaussian distributions concerning the $x/y$-direction, which is constrained by $\sum\nolimits_{l_1=1}^{L_1}{\phi _{x/y,l_1}}=\sum\nolimits_{l_2=1}^{L_2}{\phi _{x/y,l_2}}=1$; $\mathcal{N} (r_x/r_y; u_{x/y, l_1/l_2},\sigma_{x/y, l_1/l_2} ^2)$ denotes the Gaussian distribution with $r_x$ and $r_y$ being the random variables, whose means and variances are given by $u_{x/y, l_1/l_2}$ and $\sigma_{x/y, l_1/l_2} ^2$, respectively.
Thus, assuming the GMMs are mutually independent, the two-dimensional (2D) PDF for the $k$-th target can be expressed as
\begin{align}
    &p\left( \mathbf{r} \right) = p\left( x,y \right) \notag \\
    &=\sum\nolimits_{l_1=1}^{L_1}\sum\nolimits_{l_2=1}^{L_2}\phi _{x,y,l_1,l_2} \mathcal{N} (r_x;u_{x,l_1},\sigma _{x,l_1}^{2}) \notag\\
    &\qquad \qquad \qquad \qquad \qquad \times\mathcal{N} (r_y; u_{y,l_2},\sigma _{y, l_2}^{2}). \label{eq:pdf_gmm}
\end{align}
where $\phi _{x,y,l_1,l_2}\triangleq \phi _{x,l_1}^{}\phi _{y,l_2}^{}$ denotes the total weights for the 2D PDFs.

With \eqref{eq:pdf_gmm} at hand, the expression for FIM $\hat{\mathbf{J}}_{\mathbf{r}}^{}$ concerning the prior distribution $\mathbf{r}\sim p\left( \mathbf{r} \right)$ can be derived according to \eqref{eq:pdf_gmm} and can be expressed as 
\begin{align}
    &\hat{\mathbf{J}}_{\mathbf{r}}^{} =\mathbb{E} _{\mathbf{r}}\left[ \nabla _{\mathbf{r}}\ln \left( p\left( \mathbf{r} \right) \right) \nabla _{\mathbf{r}}\ln \left( p\left( \mathbf{r} \right) \right) ^{\mathrm{T}} \right] 
    \notag \\
    &=\left[ \begin{matrix}
    	\mathbb{E} _{\mathbf{r}}\left\{ \frac{\partial g\left( \mathbf{r} \right)}{\partial r_x}\frac{\partial g\left( \mathbf{r} \right)}{\partial r_x} \right\}&		\mathbb{E} _{\mathbf{r}}\left\{ \frac{\partial g\left( \mathbf{r} \right)}{\partial r_x}\frac{\partial g\left( \mathbf{r} \right)}{\partial r_y} \right\} \notag \\
    	\mathbb{E} _{\mathbf{r}}\left\{ \frac{\partial g\left( \mathbf{r} \right)}{\partial r_y}\frac{\partial g\left( \mathbf{r} \right)}{\partial r_x} \right\}&		\mathbb{E} _{\mathbf{r}}\left\{ \frac{\partial g\left( \mathbf{r} \right)}{\partial r_y}\frac{\partial g\left( \mathbf{r} \right)}{\partial r_y} \right\} \notag \\
    \end{matrix} \right]=\left[ \begin{matrix}
    	F_{xx}&		F_{xy}\\
    	F_{yx}&		F_{yy}\\
    \end{matrix} \right], \label{eq:fim_prior_distribution}
\end{align}
where $g\left( \mathbf{r} \right) \triangleq \ln \left( p\left( \mathbf{r} \right) \right) $.
Due to the position distributions being independent along the $x$- and $y$-direction, the cross terms are zero, i.e., $F_{xy}=F_{yx}=0$.
Thus, we focus on the entries on the main diagonal.
Due to the symmetry between the $x$- and $ y$-directions, we elaborate on the derivation of $F_{xx}$, while that of $F_{yy}$ can be obtained in exactly the same way.
Therefore, the following derivations can be conducted
\begin{align}
	&F_{xx} =\mathbb{E} _{\mathbf{r}}\left\{ \left( \frac{\partial g\left( \mathbf{r} \right)}{\partial r_x} \right) ^2 \right\} =\mathbb{E} _{\mathbf{r}}\left\{ \left( \frac{\partial \ln \left( p_x\left( r_x \right) \right)}{\partial r_x} \right) ^2 \right\} \notag \\
	&=\mathbb{E} _{\mathbf{r}}\left\{ \left( \frac{\partial}{\partial r_x}\left( \ln \left\{ \sum\nolimits_{l_1=1}^{L_1}{\phi _{x,l_1}^{}\mathcal{N} (r_x;u_{x,l_1}^{},\sigma _{x,l}^{2})} \right\} \right) \right) ^2 \right\} \notag \\
	&=\mathbb{E} _{\mathbf{r}}\left\{ \left( \frac{1}{p_x\left( r_x \right)}\sum\nolimits_{l_1=1}^{L_1}{\phi _{x,l_1}^{}\frac{\left( u_{x,l_1}-r_x \right)}{\sqrt{2\pi}\sigma _{x,l_1}^{3}}e^{-\frac{\left( r_x-u_{x,l_1} \right) ^2}{2\sigma _{x,l_1}^{2}}}} \right) ^2 \right\} \notag \\
	&=\mathbb{E} _{\mathbf{r}}\left\{ q_x\left( r_x \right) \right\}. 
\end{align}
As the closed-form results in the above integral are mathematically intractable, we adopt the Gauss-Hermite quadrature (GHQ) approximation to evaluate the integrals in a numerical manner, i.e., 
\begin{align}
    \mathbb{E} _{b}\left\{ f\left( b \right) \right\} =\int_{-\infty}^{+\infty}{f\left( b \right)}e^{-b^2}\mathrm{d}b\approx \sum\nolimits_{i=1}^I{w_if\left( b_i \right)}, \label{eq:GHQ}
\end{align}
where $I$ denotes the total number of terms used for this approximation, while $\{w_i\}_{i=1}^I$ and $\{b_i\}_{i=1}^I$ denote the weights and abscissa factors of Gauss-Hermite integration, respectively.
Using the additive property of the expectation operation, we can apply GHQ in \eqref{eq:GHQ} to individual Gaussian distributions in GMM \eqref{eq:pdf_gmm}.

Therefore, the following derivations can be conducted
\begin{align}
    &\mathbb{E} _{\mathbf{r}}\left\{ q_x\left( r_x \right) \right\} =\int_{\mathbb{R}}\phi _{x,l_1}^{}\sum\nolimits_{l_1=1}^{L_1}{q_x\left( r_x \right)} \notag \\ 
    &\qquad \qquad \qquad \qquad \quad \qquad \times\frac{1}{\sqrt{2\pi}\sigma _{x,l_1}^{}}e^{-\frac{\left( r_x-u_{x,l_1} \right) ^2}{2\sigma _{x,l_1}^{2}}}\mathrm{d}x \notag \\
    &=\frac{1}{\sqrt{\pi}}\sum\nolimits_{l_1=1}^{L_1}{\phi _{x,l_1}^{}\int_{\mathbb{R}}{q_x\left( \sqrt{2}\sigma _{x,l_1}^{}\tilde{x}+u_{x,l_1} \right) e^{-\tilde{x}^2}\mathrm{d}\tilde{x}}}  \notag \\
    &=\frac{1}{\sqrt{\pi}}\sum\nolimits_{i=1}^I{\sum\nolimits_{l_1=1}^{L_1}{\phi _{x,l_1}^{}w_iq_x\left( \sqrt{2}\sigma _{x,l_1}^{}\tilde{r}_{x,i}+u_{x,l_1} \right)}} . \notag
\end{align}
Using the same method, the other term $F_{yy}\left( \mathbf{r} \right)$ can be computed and approximated by a summation.
Therefore, we can derive all the diagonal entries of the FIM regarding the prior distribution.
Then, we compute the FIM regarding the observation in what follows.
To begin with, we first define $\boldsymbol{\xi} = [r_x, r_y]^{\mathrm{T}} \in \mathbb{R}^{2\times 1}$ as the unknown parameter vector for an arbitrary target.
Given the signal model described by \eqref{eq:uplink_signal}, the expression for the FIM concerning observation is specified by~\cite{shen2010fundamental}:
\begin{align} 
&\check{\mathbf{J}}_{\mathbf{r}}^{}\left( \mathbf{x} \right) \triangleq \mathbb{E} _{\mathbf{r}}\left\{ \Re \left\{ \frac{\partial \mathbf{f}_{}^{\mathrm{H}}(\mathbf{x};\mathbf{r})}{\partial \boldsymbol{\xi }}\left( \frac{1}{\sigma ^2}\mathbf{I}_M \right) \frac{\partial \mathbf{f}_{}^{}(\mathbf{x};\mathbf{r})}{\partial \boldsymbol{\xi }} \right\} \right\} \notag
\\
&=\frac{2P}{\sigma ^2}\mathbb{E} _{\mathbf{r}}\left\{ \Re \left\{ \frac{\partial \mathbf{f}_{}^{\mathrm{H}}(\mathbf{x};\mathbf{r})}{\partial \boldsymbol{\xi }}\frac{\partial \mathbf{f}_{}^{}(\mathbf{x};\mathbf{r})}{\partial \boldsymbol{\xi }} \right\} \right\} \notag
\\
&=\frac{2P}{\sigma ^2}\left[ \begin{matrix}
	\mathbb{E} _{\mathbf{r}}\left\{ J_{xx}^{}\left( \mathbf{x};\mathbf{r} \right) \right\}&		\mathbb{E} _{\mathbf{r}}\left\{ J_{xy}^{}\left( \mathbf{x};\mathbf{r} \right) \right\}\\
	\mathbb{E} _{\mathbf{r}}\left\{ J_{yx}^{}\left( \mathbf{x};\mathbf{r} \right) \right\}&		\mathbb{E} _{\mathbf{r}}\left\{ J_{yy}^{}\left( \mathbf{x};\mathbf{r} \right) \right\}\\
\end{matrix} \right] \notag \\
&=\frac{2P}{\sigma ^2}\left[ \begin{matrix}
	J_{xx}\left( \mathbf{x} \right)&		J_{xy}\left( \mathbf{x} \right)\\
	J_{yx}\left( \mathbf{x} \right)&		J_{yy}\left( \mathbf{x} \right)\\
\end{matrix} \right] \in \mathbb{R}^{2 \times 2}, \label{eq:prior_knowledge}
\end{align}
where the entries in this FIM can be calculated according to:
\begin{subequations}
    \begin{align}\label{eq:fim_entires}
       &J_{xx}\left( \mathbf{x} \right) \triangleq \notag \\
       &\quad \mathbb{E} _{\mathbf{r}}\left\{ \Re \left\{ \frac{\partial \mathbf{f}_{}^{\mathrm{H}}(\mathbf{x};\mathbf{r})}{\partial  r_x}\frac{\partial \mathbf{f}_{}^{}(\mathbf{x};\mathbf{r})}{\partial r_x} \right\} \right\} =\mathbb{E} _{\mathbf{r}}\{J_{11}(\mathbf{x};\mathbf{r})\},
        \\
        &J_{xy}\left( \mathbf{x} \right) \triangleq \notag \\ &\quad \mathbb{E} _{\mathbf{r}}\left\{ \Re \left\{ \frac{\partial \mathbf{f}_{}^{\mathrm{H}}(\mathbf{x};\mathbf{r})}{\partial  r_x}\frac{\partial \mathbf{f}_{}^{}(\mathbf{x};\mathbf{r})}{\partial  r_y} \right\} \right\} =\mathbb{E} _{\mathbf{r}}\{J_{12}(\mathbf{x};\mathbf{r})\}, 
        \\
        &J_{yx} \left( \mathbf{x} \right) \triangleq \notag \\ &\quad \mathbb{E} _{\mathbf{r}}\left\{ \Re \left\{ \frac{\partial \mathbf{f}_{}^{\mathrm{H}}(\mathbf{x};\mathbf{r})}{\partial r_y}\frac{\partial \mathbf{f}_{}^{}(\mathbf{x};\mathbf{r})}{\partial r_x} \right\} \right\} =\mathbb{E} _{\mathbf{r}}\{J_{21}(\mathbf{x};\mathbf{r})\},
        \\
        &J_{yy}\left( \mathbf{x} \right) \triangleq \notag \\ &\quad \mathbb{E} _{\mathbf{r}}\left\{ \Re \left\{ \frac{\partial \mathbf{f}_{}^{\mathrm{H}}(\mathbf{x};\mathbf{r})}{\partial r_y}\frac{\partial \mathbf{f}_{}^{}(\mathbf{x};\mathbf{r})}{\partial r_y} \right\} \right\} =\mathbb{E} _{\mathbf{r}}\{J_{22}(\mathbf{x};\mathbf{r})\}.
    \end{align}
\end{subequations}
Letting $b=\{r_x, r_y\}$ jointly represent the entries in $\boldsymbol{\xi}$, the partial derivatives in the above equations can be uniformly derived as follows:
\begin{align}
    \dot{f}_b\left( \mathbf{x};\mathbf{r} \right) =\frac{\partial \mathbf{f}_{}^{}(\mathbf{x};\mathbf{r})}{\partial b} =\mathbf{g}^{\mathrm{T}}(\mathbf{x};\mathbf{r})\frac{\partial \mathbf{h}(\mathbf{x};\mathbf{r})}{\partial b}, \notag
\end{align}
where 
\begin{align}
    &\frac{\partial \mathbf{h}(\mathbf{x};b)}{\partial b}=\eta \left[ \frac{\partial}{\partial b}\left\{ \frac{e^{-\mathrm{j}k_0r_{1}}}{r_{1}} \right\} ,...,\frac{\partial}{\partial b}\left\{ \frac{ e^{-\mathrm{j}k_0r_{M}}}{r_{M}} \right\} \right] ^{\mathrm{T}}, \notag \\
    &\frac{\partial}{\partial b}\left\{ \frac{e^{-\mathrm{j}k_0r_{m}}}{r_{m}} \right\} =\frac{\left( x_m -b \right) \left( 1+\mathrm{j}k_0r_{m} \right) e^{-\mathrm{j}k_0r_{m}}}{r_{m}^{3}}. \notag
\end{align}
As the FIM regarding the observation is parameterized by the exact position of the target, i.e, $\mathbf{r}$, which is used for the evaluation of CRB.
For BCRB, however, an additional expectation over the target position distributions, i.e., $p(\mathbf{r})$, is required to release this reliance.

More specifically, letting $i,j\in \{1,2\}$, the expectations in \eqref{eq:fim_entires} can be uniformly expressed as 
\begin{align}
    \mathbb{E} _{\mathbf{r}}\{J_{ij}(\mathbf{x};\mathbf{r})\}&=\iint_{\mathbb{R} ^{2\times 2}}{J_{ij}(\mathbf{x};\mathbf{r})p\left( \mathbf{r} \right) \mathrm{d}r_x\mathrm{d}r_y}. \label{eq:integral_form}
\end{align}
Leveraging on the two-dimensional Gauss-Hermite numerical integral method, the integrals in \eqref{eq:integral_form} can be approximated by
\begin{align}
    &\mathbb{E} _{\mathbf{r}}\{J_{ij}(\mathbf{x};\mathbf{r})\}=\frac{1}{\pi}\sum\nolimits_{l_1=1}^{L_1}{\sum\nolimits_{l_2=1}^{L_2}{\sum\nolimits_{i_1=1}^{I_1}{\sum\nolimits_{i_2=1}^{I_2}{\beta _{l_1,l_2,i_1,i_2}}}}} \notag \\
    &\qquad \times J_{ij}(\mathbf{x};\sqrt{2}\sigma _{x,l_1}^{}\tilde{x}_{i_1}+u_{x,l_1},\sqrt{2}\sigma _{y,l_2}^{}\tilde{y}_{i_2}+u_{y,l_2}), \label{eq:summation_form}
\end{align}
where the weight for each term is defined as $\beta _{l_1,l_2,i_1,i_2}\triangleq \phi _{x,l_1}^{}\phi _{y,l_2}^{}w_{i_1}w_{i_2}$.
For brevity, the derivation of \eqref{eq:summation_form} is detailed in \textbf{Appendix \ref{appendix:A}}.

Building on the previous mathematical manipulation, the BFIM can be computed according to
\begin{align}
    &\mathbf{J}_{\mathbf{r}}^{}\left( \mathbf{x} \right) =\check{\mathbf{J}}_{\mathbf{r}}^{}\left( \mathbf{x} \right) +\hat{\mathbf{J}}_{\mathbf{r}}^{} 
\notag\\
&=\left[ \begin{matrix}
	\frac{2P}{\sigma ^2}J_{xx}\left( \mathbf{x} \right) +F_{xx}&		\frac{2P}{\sigma ^2}J_{xy}\left( \mathbf{x} \right) +F_{xy}\\
	\frac{2P}{\sigma ^2}J_{yx}\left( \mathbf{x} \right) +F_{yx}&		\frac{2P}{\sigma ^2}J_{yy}\left( \mathbf{x} \right) +F_{yy}\\
\end{matrix} \right]  \notag \\
&=\left[ \begin{matrix}
	\left[ \mathbf{J}_{\mathbf{r}}^{}\left( \mathbf{x} \right) \right] _{1,1}&		\left[ \mathbf{J}_{\mathbf{r}}^{}\left( \mathbf{x} \right) \right] _{1,2}\\
	\left[ \mathbf{J}_{\mathbf{r}}^{}\left( \mathbf{x} \right) \right] _{2,1}&		\left[ \mathbf{J}_{\mathbf{r}}^{}\left( \mathbf{x} \right) \right] _{2,2}\\
\end{matrix} \right]. \label{eq:bfim}
\end{align}
Furthermore, employing the matrix inversion formula, the BCRB can be computed in a closed-form manner, which is expressed as
\begin{align}
    \mathrm{BCRB}\left( \mathbf{x} \right) =\frac{\left[ \mathbf{J}_{\mathbf{r}}^{}\left( \mathbf{x} \right) \right] _{2,2}+\left[ \mathbf{J}_{\mathbf{r}}^{}\left( \mathbf{x} \right) \right] _{1,1}}{\left[ \mathbf{J}_{\mathbf{r}}^{}\left( \mathbf{x} \right) \right] _{1,1}\left[ \mathbf{J}_{\mathbf{r}}^{}\left( \mathbf{x} \right) \right] _{2,2}-\left[ \mathbf{J}_{\mathbf{r}}^{}\left( \mathbf{x} \right) \right] _{1,2}\left[ \mathbf{J}_{\mathbf{r}}^{}\left( \mathbf{x} \right) \right] _{2,1}}. \label{eq:bcrb}
\end{align}
Here, the computation of BCRB is finished. 
\begin{remark}
     (Relationship Between BCRB and Conventional CRB) \emph{The BCRB reduces to the conventional CRB through the following two steps: i) instead of taking the expectation over the target position distribution in $\check{\mathbf{J}}_{\mathbf{r}}\left( \mathbf{x} \right)$, the FIM of the observation is evaluated at the ground-truth sensing parameters; and ii) the FIM associated with the prior distribution, $\hat{\mathbf{J}}_{\mathbf{r}}$, is omitted, as this prior information is not utilized.}
\end{remark}

\section{Single-Target Case} \label{sect:single_target}
As mentioned earlier, BCRB is the generalized lower bound for any estimator, regardless of the estimator's unbiasedness. 
Therefore, the sensing performance will be theoretically enhanced by minimizing BCRB. 
Therefore, in what follows, we consider a simple single-pinch scenario to gain insights into the interplay between PASS and sensing. 
Then, we propose an optimization framework for the general multi-pinch scenario.
  
\subsection{Single-Pinch Scenario}
In this section, we formulate the BCRB minimization problem with two assumptions and identify the optimal PA position via approximation.
Then, we provide a discussion on the optimal pinch position, revealing a counterintuitive issue in PASS sensing.

\subsubsection{Model Simplification}
To simplify the optimization problem, we make the following assumptions:
First, we assume that only the $x$-coordinate of the target needs to be estimated, with the $y$-coordinate known. 
This assumption is justified in scenarios where the target moves along a 1D track or when one spatial dimension has already been estimated using a 1D method.
Second, we assume that the $x$-coordinate follows a single Gaussian distribution, i.e., $x \sim \mathcal{N}(u_x, \sigma_x^2) $, where $u_x$ denotes the distribution centroid and $\sigma_x$ measures the uncertainty of the distribution.  
Here, $u_x$ and $\sigma_x^2$ can be seen as our prior knowledge on the distribution.
This assumption applies when the GMM has one dominant component with a significantly larger weight than the others.

With the above assumptions, the FIM related to the observation model can be simplified to one term $J_{xx}(x)$, which corresponds to the $x$-coordinate uncertainty.
Moreover, the FIM concerning the prior $x$-coordinate distribution can also be simplified via
\begin{align}
   &F_{xx} =\notag \\
   &\mathbb{E} _{\mathbf{r}}\left\{ \left( \left( \frac{1}{\sqrt{2\pi}\sigma _{x}^{}}e^{-\frac{\left( x-u_x \right) ^2}{2\sigma _{x}^{2}}} \right) ^{-1}\frac{\left( u_x-x \right)}{\sqrt{2\pi}\sigma _{x}^{3}}e^{-\frac{\left( x-u_x \right) ^2}{2\sigma _{x}^{2}}} \right) ^2 \right\} 
    \notag \\
    &=\frac{1}{\sigma _{x}^{4}}\mathbb{E} _{\mathbf{r}}\left\{ \left( u_x-x \right) ^2 \right\} =\frac{1}{\sigma _{x}^{2}},
\end{align}
which is determined solely by the prior distributions and is independent of the position of the PAs.
Building on the above, the BCRB is computed through 
\begin{align}
    \mathrm{BCRB}\left( x \right) =\left( \frac{2P}{\sigma ^2}\mathbb{E} _{\mathbf{r}}\{J_{xx}(x;r_x)\} + \sigma_x^{-2} \right) ^{-1}.
\end{align}
Further leveraging the relationship between BCRB and BFIM and excluding constants, the BCRB minimization problem can be expressed as 
\begin{subequations} 
\label{problem_single_target_single_PA}
    \begin{align}
        \min _{x} &\quad \mathrm{BCRB}\left( x \right) \label{obj:bcrb}\\
        \mathrm{s}.\mathrm{t}. &\quad  0 \le x \le x_{\max}.
    \end{align}
\end{subequations}
In \eqref{obj:bcrb}, only the Fisher information (FI) regarding the observation is determined by the PA position $x$, whereas other terms are constant.
The original BCRB minimization can be equivalently written as FI maximization, i.e., $\max _{0\le x \le x_{\max}}~\mathbb{E} _{r_x}\{J_{xx}(x;r_x)\}\triangleq F(x)$.
To solve this problem, the one-dimensional searching algorithm can be employed, which traverses the $x$ position within $[0, x_{\max}]$.
Furthermore, to address the integral term, the GHQ can be explored for the calculation of the expectation over $\mathbb{R}$.
Therefore, defining $\Delta \triangleq \sqrt{d^2 + r_y^2}$ and $r_i\triangleq\sqrt{(x-r_{x,i})^2+\Delta^2}$ with $r_{x,i}$ being the $i$-th GHQ sample, $F(x)$ can be further mathematically manipulated by the following steps:
\begin{align}
    &F(x)\approx \frac{2P}{\sqrt{\pi}\sigma ^2}\sum\nolimits_{i=1}^I{w_iJ_{xx}(x;\sqrt{2}\sigma _{x}^{}r_{x,i}}+u_x,r_y)\notag 
    \\
    &=\tilde{C}\sum_{i=1}^I{w_i\left| \frac{\left( \sqrt{2}\sigma _{x}^{}r_{x,i}+u_x-x \right) \left( 1+\mathrm{j}k_0r_i \right) e^{-\mathrm{j}\left( k_0r_i+k_{\mathrm{w}}x \right)}}{r_{i}^{3}} \right|^2} \notag
    \\
    &=\tilde{C}\sum_{i=1}^I{w_i\left| \frac{\left( \sqrt{2}\sigma _{x}^{}r_{x,i}+u_x-x \right) \left( 1+\mathrm{j}k_0r_i \right)}{r_{i}^{3}} \right|^2} \notag
    \\
    &\overset{\left( a \right)}{\approx}\tilde{C}\sum_{i=1}^I{w_i\frac{\left( \sqrt{2}\sigma _{x}^{}r_{x,i}+u_x-x \right) ^2k_{0}^{2}}{r_{i}^{4}}}. 
    \label{eq:simplified}
\end{align}
where the constant term is defined as $\tilde{C}\triangleq 2P\eta ^2/\sqrt{\pi}\sigma ^2$, and step $(a)$ is obtained by the fact that $k_{0}^{2}r_i^2\gg 1$.
It is noted that, due to the high-frequency bands adopted by PASS and the target-PA distance, which is typically on the scale of tens of meters, this approximation is both realistic and tight.
In this case, the optimal PA position $x^{\star}$ can be found by computing all stationary points using first-order optimality in \eqref{eq:simplified} and comparing these values.

\subsubsection{Discussions on the Optimal Pinch Location}
Due to the large number of stationary points on the optimization landscape, it remains challenging to gain insight into this problem.
Therefore, we further simplify this expression using empirical results obtained from simulations.
Later, the tightness of this approximation is validated to ensure its correctness.
In particular, we find that the GHQ can converge with a small number of quadrature terms, i.e., $I=3$.
Based on these empirical results and letting $C\triangleq k_0^2 \tilde{C}$, \eqref{eq:simplified} can be further simplified by keeping a handful of predominant terms:
\begin{align}
    & F(x) \approx{} \frac{2}{3}
\frac{C}{(x - u_x)^2 + 2\,\Delta^2 + \dfrac{\Delta^4}{(x - u_x)^2}} \notag \\
&\quad + \frac{1}{6}
\frac{C}{(x - (u_x + \sqrt{3}\,\sigma_x))^2 + 2\,\Delta^2 + \dfrac{\Delta^4}{(x - (u_x + \sqrt{3}\,\sigma_x))^2}} \notag \\
&\quad + \frac{1}{6}
\frac{C}{(x - (u_x - \sqrt{3}\,\sigma_x))^2 + 2\,\Delta^2 + \dfrac{\Delta^4}{(x - (u_x - \sqrt{3}\,\sigma_x))^2}}, \label{eq:ghq_simplified}
\end{align}
which aligns with the three-sigma rule in statistics.
In this case, the expression for $F(x)$ is further simplified due to the presence of a small number of critical points and can therefore be solved efficiently.
\begin{figure*}[ht]
	\centering
	\subfloat[Deterministic case with $\sigma_x^2=0.01$]{
		\includegraphics[width=0.4\linewidth]{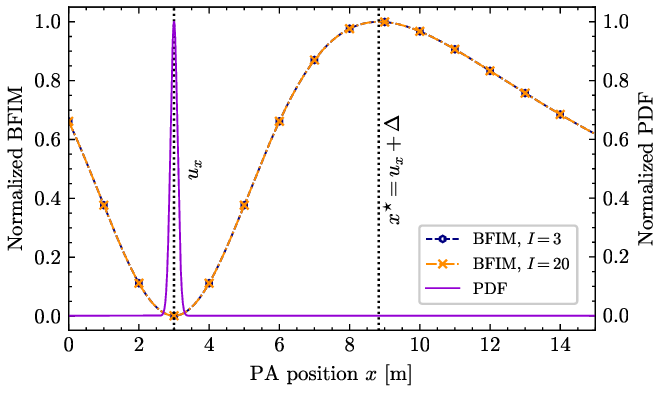} } 
	\subfloat[Random case $\sigma_x^2=3.00$]{
		\includegraphics[width=0.4\linewidth]{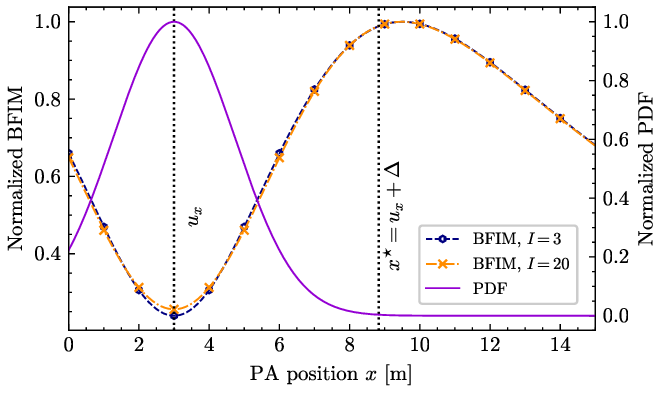}}
	\caption{Illustration of the mismatch between the distribution centroid and the sensing-sensitive centroid.} \label{fig:mismatch}
\end{figure*}
To further gain insights into the optimal placement of PA, we consider two scenarios: $\sigma_x^2 \rightarrow 0$ and $\sigma_x^2 \rightarrow \infty$.
In the first case, when $\sigma_x^2 \rightarrow 0$, FI in \eqref{eq:ghq_simplified} can be further simplified via
\begin{align}
    F(x) \approx \frac{C}{(x - u_x)^2 + 2\,\Delta^2 + \dfrac{\Delta^4}{(x - u_x)^2}}, \label{eq:PA_position_opt}
\end{align}
whose optimal value can be achieved at $x^{\star}=u_x\pm \Delta$.
In the second case, the optimal values of the second and third terms occur at $u_x \pm \sqrt{3} \sigma_x$.
However, their contributions vanish for any finite $x$ and account for only one-third of the total value.
Therefore, the optimal PA position approximately remains at  $x^{\star}\simeq u_x\pm \Delta$.
In both limits, the optimal placement is offset from the distribution mean $u_x$.
To analyze this counterintuitive observation, we present a detailed explanation of the origin of this offset. 
Before proceeding with further discussions, we provide a brief overview of FI, which will be helpful in understanding the optimal placement of PA.

\begin{remark}(Physical Interpretations of FI)\emph{
The FI quantifies the sensitivity of the measurement function (or log-likelihood function) to changes in the true value of the unknown parameter.
This sensitivity is captured by the first-order partial derivatives of the measurement function with respect to the unknown parameter.
Therefore, large FI means that even small perturbations to the value of the unknown parameter will lead to significant changes in the measurement value. 
Thus, minimizing the CRB corresponds to making the spikes in the measurement landscape as narrow and sharp as possible.}
\end{remark}

When the prior distribution of the target is considered, it is intuitive to place the PA close to the centroid of the prior distribution, i.e., the mean of the distribution.
In particular, when the density of the distribution is centered here, the Bayesian FI can be enhanced by offering a high SNR for the majority of the possible target positions.
However, this is not the case, since the distribution centroid (referred to as the PA's position that is closest to the mean of the target distribution) and the sense-sensitive centroid (referred to as the PA's position that maximizes FI) are not matched.
Recall that, according to our derivations given in \eqref{eq:PA_position_opt}, the optimal PA positions in both cases have an offset $\Delta$ to the distribution centroid $u_x$.

To make this theoretical result more understandable, we visualize this mismatch with simulations under two scenarios: 1) the distribution of target position is more deterministic, i.e., a smaller variance $\sigma_x^2=0.01$, and 2) the distribution of target position is more random, i.e., a larger variance $\sigma_x^2=3.00$.
Under the above setups, Fig. \ref{fig:mismatch} illustrates the normalized BFIM in relation to its recorded maximum value and the normalized PDF of the target position distribution.
First, we can observe that even though a small number of terms in GHQ ($I=3$) is utilized, the numerical result has already converged to that of $I=20$, which, therefore, supports the approximation used in \eqref{eq:ghq_simplified}.
Then, we can see that when PA is placed at the centroid of the distribution $u_x$, the normalized BFIM is small, which can be equivalently interpreted as a high BCRB.
However, when PA is placed with an offset of $\Delta$ to the centroid of the distribution $u_x$, the normalized BFIM is large, meaning that sensing is more sensitive at this point.
Thus, a mismatch between the distribution centroid and the sensing-sensitive centroid is visualized.
Moreover, the derived PA position $x^\star$ at $\sigma_x^2\rightarrow 0$ can still achieve a near-optimal FIM when a larger $\sigma_x^2$ is considered.
This fact shows that the optimal PA position, which minimizes the BCRB, i.e., the sensing-sensitive centroid, is deviated from the centroid of the prior $x$-direction distribution, thus revealing a mismatch between the statistical centroid and the sensing-sensitive centroid.

\begin{figure}
    \centering
    \includegraphics[width=0.8\linewidth]{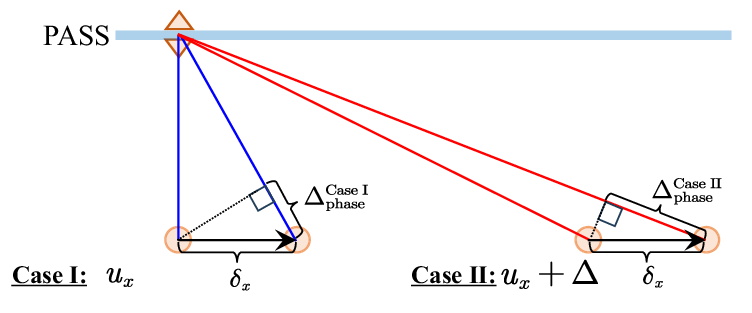}
    \caption{Illustration of Case I and Case II.} 
    \label{fig:explain_mismatch}
\end{figure}
From a geometric perspective, this mismatch is attributed to the presence of spherical waves in the near-field region.
More specifically, according to the measurement function in \eqref{eq:uplink_signal}, when placing PA at the centroid of the prior distribution, i.e., $x=u_x$, we introduce an infinitesimal perturbation to the target's position, i.e., $\delta_x$.
As a consequence, the phase changes caused by such a change in the target's position $r_x$ can be expressed as 
\begin{align}
    \Delta _{\mathrm{phase}}&=\sqrt{\delta_x ^2+\Delta ^2}\overset{\left( a \right)}{\approx}\Delta +{\delta_x ^2}/{2\Delta}, \label{eq:no_shift}
\end{align}
where step $(a)$ leverages the second-order Taylor expansion.
On the contrary, considering the second case, where the PA is placed with a shift of $\Delta$ (for convenience) to $u_x$, we have the following derivations:
\begin{align}
    \Delta _{\mathrm{phase}}=\sqrt{\left( \Delta -\delta_x \right) ^2+\Delta ^2} \overset{\left( a \right)}{\approx}\sqrt{2}\Delta-{\delta_x}/{\sqrt{2}}, \label{eq:shift}
\end{align}
where the second-order Taylor expansion is utilized at step $(a)$.
Compared \eqref{eq:shift} (Case II) with \eqref{eq:no_shift} (Case I), we can see that for the same perturbation of $\delta_x>0$, \eqref{eq:shift} is more sensitive to this change as shown in Fig. \ref{fig:explain_mismatch}, indicating FI is a decreasing function with respect to $x$ when only phase is considered.
\begin{algorithm}[t!]
\small
    \SetAlgoLined 
	\caption{Element-Wise BCRB Minimization}\label{alg:element-wise}
	\KwIn{Position distributions over the $x$- and $y$-coordinate of GMMs, Feasible region boundary $x_{\max}$, Searching resolution $\Delta_x$ \;} 
	\KwOut{Optimized PA $x$-coordinate vector $\mathbf{x}^\star$ \;}
     Initialize the searching start point $\mathbf{x}=\boldsymbol{0}_M$ and the index of PA $m=1$\;
     Set the best evaluation value via $\mathrm{f}_{\rm tar}$\;
    \Repeat{The fractional decrease of the objective value of $f(\cdot)$ falls below a predefined threshold}{
    \For{$m=1$ \KwTo $M$}
    {   
        \While{PA position satisfies $x_m^{(l)}\leq x_{\max}$ }{
        Construct completed $x$-coordinate vector via $\mathbf{x}^{(l)}=\mathbf{x}\mid_{x_m=x_m^{(l)}}$\;
        Initialize the best PA position $x_m^{\star}=0$\;
            \uIf{$\mathrm{f}(\mathbf{x}^{(l)}) \le \mathrm{f}_{\rm tar}$ and $x_m^{(l)} \in \mathcal{F}$}{
                Record the best evaluation value via $\mathrm{f}_{\rm tar} =\mathrm{f}(\mathbf{x}^{(l)}) $ \;
                  }
            Update the best $m$-th PA position via $x_m^{\star}=x_m^{(l)}$\;
            Enter the next iteration by $l=l+1$ and $x_m^{(l+1)} = x_m^{(l)} + \Delta_x$\;
    }}
    Finish the 1D searching for the $m$-th PA position $\mathbf{x}^{(l)}=\mathbf{x}\mid_{x_m=x_m^{\star}}$\;
    }
\Return{Optimized PA $x$-coordinate vector $\mathbf{x}^\star$\;}
\end{algorithm}
However, as $x$ increases, the path loss will then play a dominant role, leading to a decrease in phase sensitivity.
This explains the existence of the turning point in Fig. \ref{fig:mismatch}.

\subsection{Multi-Pinch Scenario}
Based on the single-pinch case, we can see that judicious PA repositioning is needed due to the mismatch between the centroid of the sensitive region and the centroid of the target distribution.
With this in mind, we investigate the multi-pinch scenario, where multiple PA positions necessitate optimization to minimize the BCRB of the sensing target with 2D uncertain dimensions.
Accordingly, we formulate the BCRB-based pinching beamforming optimization as follows:
\begin{subequations} \label{problem_single_target}
    \begin{align}
        \min _{\mathbf{x}} &\quad \mathrm{BCRB}\left( \mathbf{x} \right) \label{obj:crb_min_multi_pas} \\
        \mathrm{s}.\mathrm{t}. &\quad \mathbf{x} \in \mathcal{F}.
    \end{align}
\end{subequations}
In \eqref{problem_single_target}, the feasible set for PA positions is given by 
  \begin{align}
    \label{continuous_activation_constraint}
    & \mathcal{F} = \left\{ x_m \; \left| \; \begin{matrix*}[l]
      0 \le x_{m} \le x_{\max}\\
      x_{m} - x_{m-1} \ge \Delta_{\min}, \forall m
    \end{matrix*} \right. \right\},
  \end{align}
  which regulates that the spacing between adjacent PAs cannot be smaller than $\Delta_{\min}$ to avoid mutual coupling, and the feasible position region of $x_m$ for $\forall m$ is confined within the length of the waveguide.
  Due to the fractional structure and highly coupled variables in \eqref{problem_single_target}, obtaining the optimal solution to \eqref{problem_single_target} is a challenging task.
To address this challenge, we present an element-wise algorithm to solve \eqref{problem_single_target} under positional constraint, whose basic idea is to sequentially identify the positions for each antenna element through 1D searching within the feasible $x$-position set, defined by $\mathcal{F}$.
The algorithm is presented in \textbf{Algorithm \ref{alg:element-wise}} by replacing the objective function $f(\cdot)$ with the objective function \eqref{obj:crb_min_multi_pas}.
Additionally, letting the total number of iteration be $I_{\mathrm{iter}}$, the computational complexity of this element-wise algorithm is $\mathcal{O}(I_{\mathrm{iter}}JM)$, where $J=x_{\max}/\Delta_x$ denotes the total number of grids for 1D search and $\Delta_x$ controls the 1D search step size.

\section{Multiple Targets Case} \label{sect:multi_target}
In this section, we consider the multi-target scenario, where multiple targets need to be localized. Each target transmits an uplink pilot to the BS during its assigned time slots, meaning only one target is localized per time slot. 
In this case, we consider two optimization objectives: i)~total power minimization under a BCRB constraint, and ii)~min-max BCRB optimization under a power constraint.
The first objective aims to balance the device power budget and the required sensing accuracy, which is particularly relevant in power-limited systems such as battery-powered IoT devices. 
The second objective focuses on maximizing sensing accuracy within a given transmit power budget, which is crucial for high-performance applications, such as joint communication–sensing systems, where precise localization is essential.
Moreover, we examine two sensing protocols: i) PS-based sensing, where PA positions are optimized independently for each time slot, and ii) PM-based sensing, where PA positions are optimized jointly across all time slots.

\subsection{Pinch Switching-Based Sensing}
\subsubsection{Transmit Power Minimization} \label{sect:transmit_power_minimization_ps}
Under the PS-based sensing, time slots are individually optimized, as only one target accesses the BS in a given time slot.
To better illustrate scheduling across all time slots, we add a subscript $k \in \{1,2,..., K\}$ to label the targets.
Further, we define the allocated PA positions and the transmit power for individual time slots as cascaded sets, i.e., $\mathcal{X} \triangleq \left\{ \mathbf{x}_k \right\} _{k=1}^{K}$ and $\mathcal{P} \triangleq \left\{ P_k \right\} _{k=1}^{K}$.
Thus, considering the minimization of the total transmit power across all time slots, the problem can be formulated as 
\begin{subequations} \label{problem_multi_target_sum_ps}
    \begin{align}
        \min_{\mathcal{X} ,\mathcal{P}}&\quad  \sum\nolimits_{k=1}^K{P_k} \label{objective:min_power}\\
        \,\,\mathrm{s}.\mathrm{t}. &\quad \mathrm{BCRB}_k\left( \mathbf{x}_k,P_k \right) \le \Gamma _{\mathrm{sen}},\forall k. \\
        &\quad \mathbf{x}_k\in \mathcal{F} ,\quad \forall~\mathbf{x}_k\in \mathcal{X}. \\ 
         &\quad P_k\ge 0,\quad \forall~k, 
    \end{align}
\end{subequations}
where $\Gamma_{\rm sen}$ is the BCRB threshold set for all sensing targets.
Following the basic idea of element-wise optimizations, we first derive the optimal power allocation policy.
In this case, the original problem can be converted into an equivalent form determined solely by PA positions.
Consequently, the PA position optimization is conducted via a 1D search.
Denoting $\check{\mathbf{J}}_{\mathbf{r}_k}^{}\left( \mathbf{x}_k \right) \triangleq \frac{2P_k}{\sigma ^2}\tilde{\mathbf{J}}_{\mathbf{r}_k}^{}\left( \mathbf{x}_k \right) $ , the subproblem with respect to $\mathcal{P}$ can be formulated as 
\begin{subequations}
    \begin{align}
        \min_{\mathcal{P}}&\quad  \sum\nolimits_{k=1}^K{P_k} \\
        \,\,\mathrm{s}.\mathrm{t}. &\quad \mathrm{tr}\left\{ \left( \frac{2P_k}{\sigma ^2}\tilde{\mathbf{J}}_{\mathbf{r}}^{}\left( \mathbf{x}_k \right) +\hat{\mathbf{J}}_{\mathbf{r}_k}^{} \right) ^{-1} \right\} \le \Gamma _{\mathrm{sen}},~\forall k.  \\
        &\quad P_k\ge 0,~\forall~k, \label{constraint:transmit_power_ge_0}
    \end{align}
\end{subequations}
where $\hat{\mathbf{J}}_{\mathbf{r}_k}^{}$ denotes the FIM regarding the prior position distribution of the $k$-th target.
Thus, to reduce the number of multipliers, we ignore constraint \eqref{constraint:transmit_power_ge_0} for now,
Thus, the Lagrangian function can be written as 
\begin{align}
    &\mathcal{L} =\sum\nolimits_{k=1}^K{P_k}\notag \\
    &+ \sum\nolimits_{k=1}^K{\lambda _k\left( \mathrm{tr}\left\{ \left( \frac{2P_k}{\sigma ^2}\tilde{\mathbf{J}}_{\mathbf{r}}^{}\left( \mathbf{x}_{k}^{} \right) +\hat{\mathbf{J}}_{\mathbf{r},k}^{} \right) ^{-1} \right\} - \Gamma _{\mathrm{sen}} \right)},
\end{align}
where $\lambda_k \ge 0 $ are the Lagrangian multipliers.
Then, the KKT condition can be listed as follows:
\begin{align}
    \begin{cases}
	\frac{\partial \mathcal{L}}{\partial P_k}=1-\lambda _k\left( \frac{2P_k}{\sigma ^2}\tilde{\mathbf{J}}_{\mathbf{r}_k}^{}\left( \mathbf{x}_{k}^{} \right) +\hat{\mathbf{J}}_{\mathbf{r}_k}^{} \right) ^{-2}\frac{2}{\sigma ^2}\tilde{\mathbf{J}}_{\mathbf{r}_k}^{}\left( \mathbf{x}_{k}^{} \right) =0,\\
	\frac{\partial \mathcal{L}}{\partial \lambda _k}=\mathrm{tr}\left\{ \left( \frac{2P_k}{\sigma ^2}\tilde{\mathbf{J}}_{\mathbf{r}_k}^{}\left( \mathbf{x}_{k}^{} \right) +\hat{\mathbf{J}}_{\mathbf{r}_k}^{} \right) ^{-1} \right\} -\Gamma _{\mathrm{sen}}=0,\\
	\lambda _k\left( \mathrm{tr}\left\{ \left( \frac{2P_k}{\sigma ^2}\tilde{\mathbf{J}}_{\mathbf{r}_k}^{}\left( \mathbf{x}_{k}^{} \right) +\hat{\mathbf{J}}_{\mathbf{r}_k}^{} \right) ^{-1} \right\} -\Gamma _{\mathrm{sen}} \right) =0.\\ 
\end{cases}\notag
\end{align}
Based on the KKT condition, the optimal transmit power denoted by $P_k^\star$ is given by
\begin{align}
    \mathrm{tr}\left\{ \left( 2P_k/\sigma ^2\tilde{\mathbf{J}}_{\mathbf{r}_k}^{}\left( \mathbf{x}_{k}^{} \right) +\hat{\mathbf{J}}_{\mathbf{r}_k}^{} \right) ^{-1} \right\} =\Gamma _{\mathrm{sen}}.
\end{align}
Therefore, the optimal policy allocation is specified by
\begin{align}
    P_k(\mathbf{x}_k)= \frac{-A_{2,k}+\sqrt{A_{2,k}^{2}-4A_{1,k}A_{3,k}}}{2A_{1,k}}, \label{eq:optimal_power_switching_min_power}
\end{align}
where $\alpha \triangleq 2 / \sigma^2$ and the rest terms are defined as follows:
\begin{align}
    &A_{1,k}\triangleq \alpha ^2\Gamma _{\mathrm{sen}}\left( J_{xx}\left( \mathbf{x}_k \right) J_{yy}\left( \mathbf{x}_k \right) -J_{xy}\left( \mathbf{x}_k \right) J_{yx}\left( \mathbf{x}_k \right) \right), \notag \\
    &A_{2,k}\triangleq \alpha \Gamma _{\mathrm{sen}}\left( J_{xx}\left( \mathbf{x}_k \right) F_{yy}\left( \mathbf{r}_k \right) +F_{xx}\left( \mathbf{r}_k \right) J_{yy}\left( \mathbf{x}_k \right) \right) \notag \\
    &\qquad -\alpha \left( J_{yy}\left( \mathbf{x}_k \right) +J_{xx}\left( \mathbf{x}_k \right) \right), \notag\\
    &A_{3,k}\triangleq \Gamma _{\mathrm{sen}}F_{xx}\left( \mathbf{r}_k \right) F_{yy}\left( \mathbf{r}_k \right) -\left( F_{yy}\left( \mathbf{r}_k \right) +F_{xx}\left( \mathbf{r}_k \right) \right). \notag  
\end{align}
For a consistent presentation, the derivation of the above equation is presented in Appendix \ref{appendix_B}.
Then, plugging $P_k(\mathbf{x}_k)$ back to \eqref{objective:min_power}, the resultant $k$-th subproblem can be written as 
\begin{align}
    \mathbf{x}_{k}^{\star}=\mathop {\mathrm{argmin}
    } \nolimits_{\mathbf{x}_{k}^{}\in \mathcal{F}}\,\,P_k\left( \mathbf{x}_k \right). \label{problem_x_position}
\end{align}
Therefore, the PA position set can be constructed by $\mathcal{X}=\{\mathbf{x}_k\}_{k=1}^{K}$.
To solve \eqref{problem_x_position}, the element-wise algorithm can be utilized by replacing $f(\mathbf{x}_k)$ with $P_k(\mathbf{x}_k)$.
This approach is summarized in \textbf{Algorithm \ref{alg:element-wise}}.
The computational complexity of this algorithm is $\mathcal{O}(I_{\mathrm{iter}}JKM)$.

\subsubsection{Min-Max BCRB} \label{sect:min_max_bcrb_ps}
Then, we consider the min-max BCRB objective, which can be formulated as
\begin{subequations} \label{problem_multi_target_min_max}
    \begin{align}
        \min_{\mathcal{X} ,\mathcal{P}} &\max  \left\{ \left\{ \mathrm{BCRB}_k\left( \mathbf{x}_k,P_k \right) \right\} _{k=1}^{K} \right\} \\
        \mathrm{s}.\mathrm{t}. &\quad \mathbf{x}_k \in \mathcal{F}, \quad \forall \mathbf{x}_k \in \mathcal{X}. \\
        &\quad \sum\nolimits_{k=1}^K{P_k}=P_{\max}. \label{constraint:transmit_power}
    \end{align}
\end{subequations}
where $P_{\max}$ denotes the total power budget of the targets.
First, we consider the optimization of power location $\mathcal{P}$, fixing the PA positions $\mathcal{X}$.
Furthermore, by introducing an auxiliary variable $u$, the original problem in \eqref{problem_multi_target_min_max} can be recast as
\begin{subequations} \label{problem_multi_target_min_max_u}
    \begin{align}
        \min_{u, \mathcal{P}} &\quad u  \\
        \mathrm{s}.\mathrm{t}. &\quad u \ge \mathrm{tr}\left\{ \left( \frac{2P_k}{\sigma ^2}\tilde{\mathbf{J}}_{\mathbf{r}_k}^{}\left( \mathbf{x}_{k} \right) +\hat{\mathbf{J}}_{\mathbf{r}_k}^{} \right) ^{-1} \right\},~ \forall k. \\
       &\quad \eqref{constraint:transmit_power}. \notag 
    \end{align}
\end{subequations}
Given the convexity of this problem, we resort to the KKT condition to drive the optimal power allocation policy.
The Lagrangian function of problem \eqref{problem_multi_target_min_max_u} is given by
\begin{align}
    \mathcal{L} &=u+\sum\nolimits_{k=1}^K{\lambda _k\left( \mathrm{tr}\left\{ \left( \frac{2P_k}{\sigma ^2}\tilde{\mathbf{J}}_{\mathbf{r}_k}^{}\left( \mathbf{x}_{k} \right) +\hat{\mathbf{J}}_{\mathbf{r}_k}^{} \right) ^{-1} \right\} -u \right)} \notag \\
    &+\lambda _0\left( \sum\nolimits_{k=1}^K{P_k-P_{\max}} \right) ,
\end{align}
where $\lambda _k \ge 0$ for $\forall k$ are the Lagrangian multipliers.
Accordingly, the KKT condition is derived as 
\begin{align}
\begin{cases}
	\frac{\partial \mathcal{L}}{\partial u}=1-\sum\nolimits_{k=1}^K{\lambda _k}=0,\\
	\frac{\partial \mathcal{L}}{\partial P_k}=-\lambda _k\mathrm{tr}\left\{ \left( \frac{2P_k}{\sigma ^2}\tilde{\mathbf{J}}_{\mathbf{r}_k}^{}\left( \mathbf{x}_k \right) +\hat{\mathbf{J}}_{\mathbf{r}_k}^{} \right) ^{-2}\frac{2}{\sigma ^2}\tilde{\mathbf{J}}_{\mathbf{r}_k}^{}\left( \mathbf{x}_k \right) \right\}\\
	\qquad +\lambda _0=0, ~\mathrm{for}~\forall~k, \\
	\lambda _k\left( \mathrm{tr}\left\{ \left( \frac{2P_k}{\sigma ^2}\tilde{\mathbf{J}}_{\mathbf{r}_k}^{}\left( \mathbf{x}_k \right) +\hat{\mathbf{J}}_{\mathbf{r}_k}^{} \right) ^{-1} \right\} -u \right) =0,~ \mathrm{for}~\forall k, \\
	\lambda _0\left( \sum\nolimits_{k=1}^K{P_k-P_{\max}} \right) =0. \\
\end{cases} \label{eq:KKT_min_max}
\end{align}
From the above conditions, the optimal values denoted by $u^\star$ and $\{ P_k^\star \}$ satisfies the following condition:
\begin{align}
    \begin{cases}
	\mathrm{tr}\left\{ \left( \frac{2P_{k}^{\star}}{\sigma ^2}\tilde{\mathbf{J}}_{\mathbf{r}_k}^{}\left( \mathbf{x}_{k} \right) +\hat{\mathbf{J}}_{\mathbf{r}_k}^{} \right) ^{-1} \right\} -u^{\star}=0,~ \mathrm{for}~\forall k, \\
	\sum\nolimits_{k=1}^K{P_{k}^{\star}-P_{\max}}=0. \\
\end{cases} 
\end{align}
According to the closed-form results presented in \eqref{eq:bfim} and \eqref{eq:bcrb}, $u^\star$ can be computed by solving the following equation:
\begin{align}
    \sum\nolimits_{k=1}^K{\frac{-A_{2,k}\left( u \right) +\sqrt{A_{2,k}^{2}\left( u \right) -4A_{1,k}\left( u \right) A_{3,k}\left( u \right)}}{2A_{1,k}\left( u \right)}=P_{\max}}, \label{eq:optimal_u}
\end{align}
where 
\begin{subequations}
    \begin{align}
    &A_{1,k}\left( u \right) \triangleq \alpha ^2u\left( J_{xx}\left( \mathbf{x}_k \right) J_{yy}\left( \mathbf{x}_k \right) -J_{xy}\left( \mathbf{x}_k \right) J_{yx}\left( \mathbf{x}_k \right) \right), \notag \\
    &A_{2,k}\left( u \right) \triangleq \alpha u\left( J_{xx}\left( \mathbf{x}_k \right) F_{yy,k}+F_{xx,k}J_{yy}\left( \mathbf{x}_k \right) \right) \notag \\
    &\qquad \qquad \qquad -\alpha \left( J_{yy}\left( \mathbf{x}_k \right) +J_{xx}\left( \mathbf{x}_k \right) \right), \notag \\
    &A_{3,k}\left( u \right) \triangleq uF_{xx,k}F_{yy,k}-\left( F_{yy,k}+F_{xx,k} \right) . \notag 
    \end{align}
\end{subequations}
The derivation of \eqref{eq:optimal_u} is similar to Appendix \ref{appendix_B}, and, therefore, is omitted here for brevity.
Note that the left-hand side (LHS) in \eqref{eq:optimal_u} is a monotonically decreasing function.
Therefore, $u^\star$ can be effectively solved using bisection methods.
In this case, the optimal power allocation policy can be obtained for a given $\{\mathbf{x}_k\}_{k=1}^{K}$.
Then, by adopting alternative optimization (AO), problem \eqref{problem_multi_target_min_max} can be solved.

To reduce the computational cost for AO, we derive the approximated solution at high SNR, i.e., large $P_k/\sigma^2$.
In this case, the third equation in \eqref{eq:KKT_min_max} can be simplified by
\begin{align}
    &\lambda _k\left( \mathrm{tr}\left\{ \left( \frac{2P_k}{\sigma ^2}\tilde{\mathbf{J}}_{\mathbf{r}_k}^{}\left( \mathbf{x}_k \right) +\hat{\mathbf{J}}_{\mathbf{r}_k}^{} \right) ^{-1} \right\} -u \right) \notag \\ 
    &\qquad \simeq \lambda _k\left( \mathrm{tr}\left\{ \left( \frac{2P_k}{\sigma ^2}\tilde{\mathbf{J}}_{\mathbf{r}_k}^{}\left( \mathbf{x}_k \right) \right) ^{-1} \right\} -u \right) =0.
\end{align}
According to the first stationary condition in \eqref{eq:KKT_min_max}, we have $\lambda_k \neq 0$ for $\forall k$,
Thus, the optimal auxiliary variable is given 
\begin{align}
    u^\star(\mathbf{x}_k)=\frac{\sigma ^2}{2P_{\max}}\sum\nolimits_{k=1}^K{\mathrm{tr}\left\{ \left( \tilde{\mathbf{J}}_{\mathbf{r}_k}^{}\left( \mathbf{x}_k \right) \right) ^{-1} \right\}}.
\end{align}
Therefore, the optimal power is given by
\begin{align}
    P_{k}^{\star}\left( \mathbf{x}_k \right) =\frac{P_{\max}\mathrm{tr}\left\{ \left( \tilde{\mathbf{J}}_{\mathbf{r}_k}^{}\left( \mathbf{x}_k \right) \right) ^{-1} \right\}}{\sum\nolimits_{k=1}^K{\mathrm{tr}\left\{ \left( \tilde{\mathbf{J}}_{\mathbf{r}_k}^{}\left( \mathbf{x}_k \right) \right) ^{-1} \right\}}}. \label{eq:optimal_power}
\end{align}
By plugging \eqref{eq:optimal_power} into the objective of \eqref{problem_multi_target_min_max}, the remaining problem can be written as
\begin{align}
    \mathbf{x}_{k}^{\star}=\mathop {\mathrm{argmin}} \nolimits_{\mathbf{x}_{k}^{}\in \mathcal{F}}\,\,u^{\star}(\mathbf{x}_k),
\end{align}
which can be solved via element-wise \textbf{Algorithm \ref{alg:element-wise}}.
The computational complexity for solving problem \eqref{problem_multi_target_min_max} can be computed by $\mathcal{O}(I_{\mathrm{iter}}\log(N)JKM)$ with $N$ being the number of discrete points used in the obtainment of the optimal power allocation policy.
Further, under the high SNR condition, the computational complexity can be reduced to $\mathcal{O}(I_{\mathrm{iter}}JKM)$, since the closed-form power for each time slot can be derived.

\subsection{Pinch Multiplexing-Based Sensing}
\subsubsection{Transmit Power Minimization} \label{sect:transmit_power_minimization_pm}
In this case, a single PA position vector is used for all time slots.
Therefore, the transmit power minimization problem can be formulated as
\begin{subequations} \label{problem_multi_target_sum_pm}
    \begin{align}
        \min_{\mathbf{x} ,\mathcal{P}}&\quad  \sum\nolimits_{k=1}^K{P_k} \label{objective:min_power_multiplexing}\\
        \,\,\mathrm{s}.\mathrm{t}. &\quad \mathrm{BCRB}_k\left( \mathbf{x}, P_k \right) \le \Gamma _{\mathrm{sen}},~\forall k. \\
        &\quad \mathbf{x} \in \mathcal{F}, \quad P_k\ge 0,~\forall~k. 
    \end{align}
\end{subequations}
To solve this problem, a similar method used in solving problem \eqref{problem_multi_target_sum_ps} can be utilized.
Particularly, the optimal transmit power $P_k^\star(\mathbf{x})$ can be solved by exploring the KKT condition of problem \eqref{problem_multi_target_sum_pm}.
The expression of $P_k^\star(\mathbf{x})$ aligns with \eqref{eq:optimal_power_switching_min_power}.
Finally, plugging in $P_k^\star(\mathbf{x})$, the remaining position optimization problem can be specified by
\begin{align}
   \mathbf{x}_{}^{\star}=\mathop {\mathrm{argmin}} \nolimits_{\mathbf{x}_{}^{}\in \mathcal{F}}\,\,\sum \nolimits_{k=1}^K{P_k\left( \mathbf{x} \right)}.\label{problem_x_position_multiplexing}
\end{align}
This problem can be solved via \textbf{Algorithm \ref{alg:element-wise}}.
It is worth noting that, since all time slots share the same PA layout, the computational complexity is reduced from $\mathcal{O}(I_{\mathrm{iter}}JKM)$ to $\mathcal{O}(I_{\mathrm{iter}}JM)$.

\subsubsection{Min-Max BCRB}\label{sect:min_max_bcrb_pm}
Concerning the min-max BCRB, the problem in the PM case can be cast as
\begin{subequations} \label{problem_multi_target_min_max_multiplexing}
    \begin{align}
        \min_{\mathbf{x} ,\mathcal{P}} &~~\max \left\{ \left\{ \mathrm{BCRB}_k\left( \mathbf{x},P_k \right) \right\} _{k=1}^{K} \right\}   \\
        \mathrm{s}.\mathrm{t}. &\quad \mathbf{x} \in \mathcal{F}, \quad \sum\nolimits_{k=1}^K{P_k}=P_{\max}.
    \end{align}
\end{subequations}
This problem can be solved in the same manner as problem \eqref{problem_multi_target_min_max} by exploring the KKT conditions.
Further, considering the high SNR condition, the final problem is given by
\begin{align}
    \mathbf{x}^{\star}=\mathop {\mathrm{argmin}} \nolimits_{\mathbf{x}^{}\in \mathcal{F}}\,\,u^{\star}(\mathbf{x}).
\end{align}
This problem can be solved with the element-wise method in \textbf{Algorithm \ref{alg:element-wise}} with a computational complexity of $\mathcal{O}(I_{\mathrm{iter}}JM)$.

\section{Numerical Results} \label{sect:simulation_results}
In this section, numerical results are presented to validate the effectiveness of the proposed methods.
Unless otherwise specified, the system parameters are set as follows:
For the physical layer, the carrier frequency is  $f=28
~\mathrm{GHz}$, the maximum transmit power is $P_{\max} = 10 ~\mathrm{dBm}$, and the noise power is $\tilde{\sigma}^2=-90~\mathrm{dBm}$. 
For the PASS layout, the waveguide extends along the $x$-axis with a side length of $x_{\max}=10~{\rm m}$ with $d=3~\mathrm{m}$.
The number of PAs is $M=5$, each adjustable within the range $[0~\mathrm{m}, x_{\max}~\mathrm{m}]$.
For the 1D search algorithm, the step size is set to $\Delta_x = 0.1~\mathrm{m}$, while the number of GH points is set to $L=150$ for each of the Gaussian distributions.
For the sensing targets, the number of targets is $K=5$, whose positions follow mixed Gaussian distributions with $L_1=L_2=1$ along both the $x$- and $y$-axes.
The means of these distributions are uniformly drawn from the region $\mathcal{R} \triangleq[-5~\mathrm{m}, x_{\max}+5~\mathrm{m}] \times [-15~\mathrm{m}, +15~\mathrm{m}]$, while the variances are uniformly drawn from $[0.01~\mathrm{m}^2, 0.5~\mathrm{m}^2]$.
To validate the effectiveness of the proposed algorithms and highlight the superiority of PASS, we introduced the following benchmarks.
\begin{figure}
    \centering
    \includegraphics[width=0.85\linewidth]{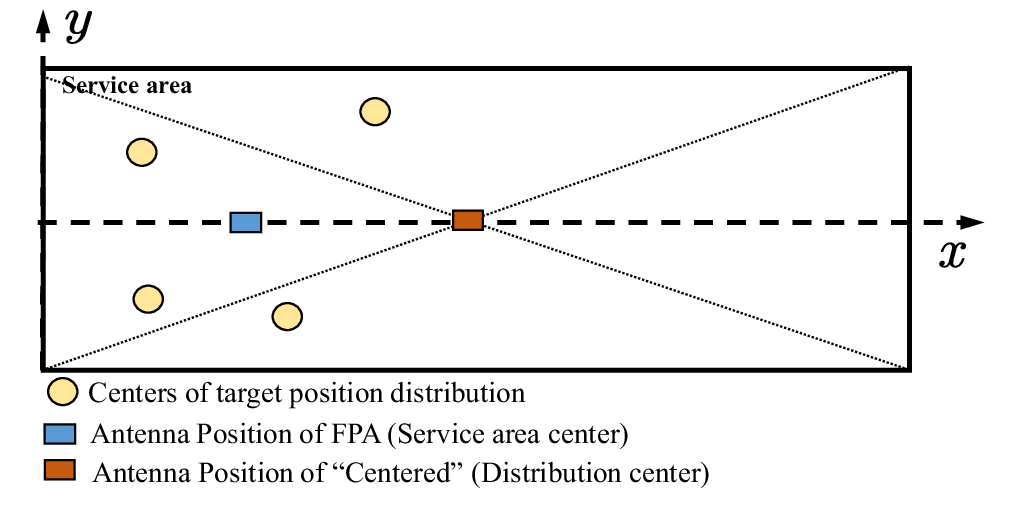}
    \caption{Illustration of the difference of antenna placements in the ``Uniform" and the ``FPA" benchmarks.}
    \label{fig:illustration}
\end{figure}
\begin{itemize}
    \item \textbf{Uniform PASS (Uniform)}: In this benchmark, the PAs on PASS are evenly spaced with an interval of $x_{\max} / (M-1)$, implying that the antennas are uniformly placed and occupy the entire waveguide.
    This benchmark presents a heuristic placement of PAs without dedicated designs.

    \item \textbf{Centered PASS (Centered)}: In this benchmark, the PAs on waveguides are packed with a uniform spacing of $\lambda/2$ to form a subarray.
    The subarray is positioned such that its midpoint aligns with the centroid of the weighted means of the targets.
    This benchmark presents another heuristic placement of PAs aiming at minimizing the average pathloss over the entire distribution of targets.

    \item \textbf{Conventional Fixed Position Array (FPA)}: 
    In order to show the superiority of PASS over conventional antennas, we introduce a conventional antenna benchmark. 
    In this benchmark, FPA is utilized, whose antenna elements are fixed and separated by half-wavelength.
    This array is placed at the center of the distribution region $\mathcal{R}$.
    For fairness, its phase optimization is constrained by the unit-modulus condition.
    Phase shifts are updated using the block coordinate descent (BCD) algorithm \cite{wang2025beamfocusing}.
    It is essential to note that, compared to the ``Uniform" benchmark, FPA is positioned at the center of the service area, rather than the distribution center, which is illustrated by Fig. \ref{fig:illustration} \footnote{This is only an illustration figure to demonstrate the difference between ``Uniform" and ``FPA". The results in the remaining figures are obtained through Monte Carlo simulations rather than a fixed setup. }.
\end{itemize}

\begin{figure}[t!]
    \centering
    \includegraphics[width=0.8\linewidth]{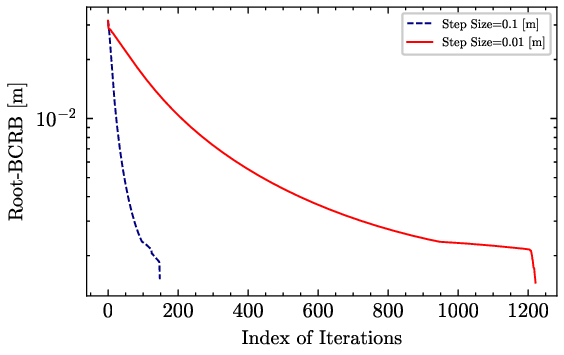}
    \caption{Illustration of the convergence behavior of 1D search under different step sizes.}
    \label{fig:convergence_1d_search}
\end{figure}
\begin{figure}[t!]
    \centering
    \includegraphics[width=0.8\linewidth]{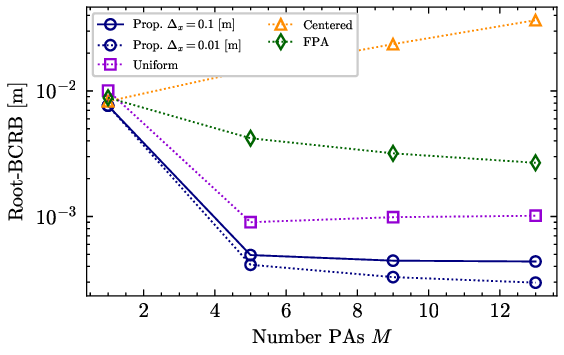}
    \caption{Illustration of root BCRB versus the number of PAs.}
    \label{fig:rbcrb_vs_num_of_pas}
\end{figure}
Fig. \ref{fig:convergence_1d_search} illustrates the convergence behavior of the 1D search for BCRB minimization.
The search process unfolds with a consistently decreasing objective function, proving the effectiveness of the proposed algorithm.
Moreover, when finer grids, i.e., smaller step sizes, are adopted, a lower BCRB can be reached yet with more iterations.
Hence, the chosen step size of $\Delta_x=0.1~\mathrm{m}$ can strike a good balance between performance and time consumption.

Fig. \ref{fig:rbcrb_vs_num_of_pas} demonstrates the value of root BCRB as a function of the number of PAs.
It can be observed that as the number of PAs increases, the value of BCRB decreases correspondingly.
However, such a decrease reaches a flat region quickly, indicating that continuously increasing $M$ can only offer minimizing returns but increased search complexity.
Therefore, the chosen PA number, i.e., $M=5$, can achieve a good tradeoff between performance and complexity.
Furthermore, the benefit brought by finer search grids is enlarged as $M$ increases.
For the benchmark algorithms, it can be seen that the PASS-based solution with heuristic PA placement, i.e., ``Uniform" and ``Centered", cannot achieve a decent performance, indicating that judicious PA position optimization is required.
Moreover, the superiority of PASS is also proved by its performance gain over FPAs.

\begin{figure}[t!]
    \centering
    \includegraphics[width=0.8\linewidth]{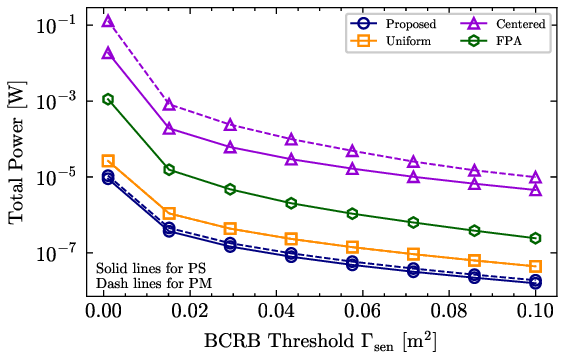}
    \caption{Illustration of total transmit power $\sum\nolimits_{k=1}^K{P_k}$ versus BCRB threshold $\Gamma_{\rm sen}$.}
    \label{fig:trans_power_vs_bcrb_threshold}
\end{figure}
Fig. \ref{fig:trans_power_vs_bcrb_threshold} illustrates the variation of the total transmit power, i.e., $\sum\nolimits_{k=1}^K{P_k}$, versus the targeted sensing threshold, i.e., $\Gamma_{\mathrm{sen}}$, under the PS-based sensing described in Section \ref{sect:transmit_power_minimization_ps} for the PS protocol and Section \ref{sect:transmit_power_minimization_pm} for the PM protocol.
This figure corresponds to the transmit power minimization problem constrained by the fixed sensing threshold, i.e., problems \eqref{problem_multi_target_sum_ps} and \eqref{problem_multi_target_sum_pm}.
As the sensing threshold $\Gamma_{\rm sen}$ increases, the total power decreases, indicating that a less stringent sensing threshold allows for lower total transmit power at all the targets.
Furthermore, owing to the high degree of optimization provided by separately optimizing PA positions for each time slot, the PS protocol consistently outperforms its PM counterpart.
In addition, under the same sensing threshold, PASS requires less total transmit power compared with the ``FPA" benchmark.
It is also noted that the curves representing the ``Uniform" benchmark overlap for the PS and PM protocols, since the PAs are uniformly placed along the entire waveguide and are not optimized.

\begin{figure}[t!]
    \centering
    \includegraphics[width=0.8\linewidth]{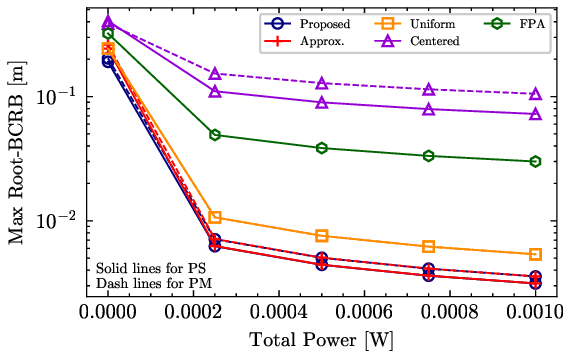}
    \caption{Illustration of maximum root BCRB versus total power budget $P_{\max}$.}
    \label{fig:max_bcrb_vs_total_power_budget}
\end{figure}
Fig. \ref{fig:max_bcrb_vs_total_power_budget} illustrates the variances of the maximum root BCRB among all targets, i.e., $\max \left\{ \sqrt{\mathrm{BCRB}}_1,...,\sqrt{\mathrm{BCRB}}_K \right\}$, versus the total power budget, i.e., $P_{\max }$, under the PS protocol in Section \ref{sect:min_max_bcrb_ps} and the PM protocol in Section \ref{sect:min_max_bcrb_pm}.
This figure corresponds to the min-max BCRB problem with a fixed total power budget.
As more power becomes available at the targets, the maximum BCRB across all targets decreases, resulting from enhanced uplink SNRs at PASS.
Compared with the ``FPA" benchmark, PASS provides significant performance gains due to its reconfigurability.
The high-SNR approximations devised for the min-max BCRB problems are also evaluated in this figure, which are labeled as ``Approx.".
The results suggest that in high SNR regions, approximation methods that intentionally omit the prior FIM can produce reliable results with lower computational complexity, since the bisection search can be discarded.
It is observed that as the total power budget $P_{\max}$ increases, i.e., a high SNR, the gap between the exact values and the approximated results diminishes, thus validating these approximations.
The overlapping curves for the ``Uniform” benchmark are due to the fixed PA positions.

\begin{figure}[t!]
    \centering
    \includegraphics[width=0.95\linewidth]{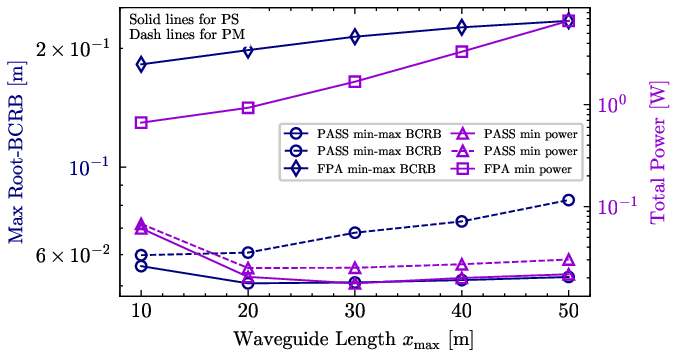}
    \caption{Illustration of maximum root BCRB and total transmit power versus the side length of the waveguide.
    The number of sensing targets, the total power budget, and the sensing threshold are fixed as $K=5$, $P_{\max}=10~\mathrm{dBm}$, and $\Gamma_{\rm sen}=0.02~\mathrm{m}^2$.}
    \label{fig:bcrb_and_min_power_vs_xmax}
\end{figure}

Fig. \ref{fig:bcrb_and_min_power_vs_xmax} jointly shows the maximum BCRB and the total transmit power as functions of waveguide length $x_{\max}$.
It is essential to note that the curves related to PASS are derived from the proposed algorithms. 
For comparison, the ``FPA" benchmark is also presented to underpin the superiority of PASS.
For the min-max BCRB problem, as the waveguide length $x_{\max}$ increases, the maximum BCRB among all targets decreases under the PS protocol but increases under the PM protocol.
This is because, as $x_{\max}$ grows, the possible positions of targets expand dramatically.
Due to its limited optimization dimension, the PM protocol cannot effectively address this more challenging scenario, resulting in performance degradation.
In contrast, owing to its enhanced optimization dimension, the PS protocol can fully harness the benefits of PASS's enhanced positional flexibility, yielding better performance.
For the total transmit power minimization problem, the total transmit power is less susceptible to changes in $x_{\max}$, resulting in a slight performance gain as $x_{\max}$ increases.
However, a growing gap emerges between the performances of the PM and PS protocols.
Under both protocols, the ``FPA" benchmark does not offer decent performance.
More importantly, the upward trend of the ``FPA" benchmark indicates that the limited flexibility of the conventional array results in poor scalability as the target distribution area increases.

\begin{figure}[t!]
    \centering
    \includegraphics[width=0.95\linewidth]{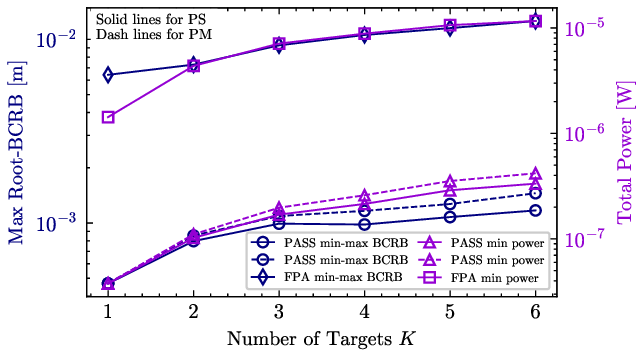}
    \caption{Illustration of maximum root BCRB and total transmit power versus the number of targets. The number of sensing targets, the total power budget, and the sensing threshold are fixed as $P_{\max}=10~\mathrm{dBm}$, and $\Gamma_{\rm sen}=0.02~\mathrm{m}^2$.}
    \label{fig:bcrb_and_min_power_vs_num_targets}
\end{figure}
Fig. \ref{fig:bcrb_and_min_power_vs_num_targets} jointly shows the maximum root BCRB and the total transmit power as functions of the number of sensing targets, i.e., $K$.
It can be observed that as the number of sensing targets increases, both the PM and PS protocols suffer from an increasing max BCRB and total transmit power.
Compared with the results in Fig. \ref{fig:bcrb_and_min_power_vs_num_targets}, increasing the optimization dimension in the PS protocol does not improve the results for either optimization setup, resulting in upward trends.
Nevertheless, the PS protocol still provides better performance than its PM counterpart, and this performance advantage becomes more pronounced as $K$ increases.

\section{Conclusions} \label{sect:conclusion}
This paper studied an uplink sensing framework for PASS.
For the single-target scenario, we first considered a single-PA case to gain insights, and then devised an element-wise algorithm for the multi-PA case.
Based on the derivation of the optimal PA position in the single-PA case, we revealed a mismatch between the centroid of the target distribution and the centroid of the highly sensitive region, which constitutes a unique feature of PASS-based sensing systems.
For the multi-target scenario, we devised two uplink accessing protocols, namely PM and PS.
Under both protocols, we investigated the total transmit power minimization and min-max BCRB problems, which were reformulated into equivalent forms parameterized solely by PA positions using the KKT conditions, and solved via element-wise algorithms.
Numerical results demonstrated that i)~the proposed algorithms effectively solve the formulated problems, outperforming heuristic baselines; ii)~with greater flexibility, PASS achieves significant gains over conventional FPA; and iii)~the PS protocol consistently outperforms the PM protocol, albeit at the cost of higher complexity.

\appendices
\section{Derivations of the BCRB in \eqref{eq:summation_form}}\label{appendix:A}
With the PDF for GMM being specified by \eqref{eq:pdf_gmm}, we have the following derivations:
    \begin{align}
        &\iint_{\mathbb{R} ^{2\times 2}}{J_{ij}(\mathbf{x};\mathbf{r})p\left( x,y \right) \mathrm{d}x\mathrm{d}y} \notag \\
    &=\iint_{\mathbb{R} ^{2\times 2}}J_{ij}(\mathbf{x};\mathbf{r})\sum\nolimits_{l_1=1}^{L_1}\sum\nolimits_{l_2=1}^{L_2}\phi _{x,l_1}^{}\phi _{y,l_2}^{}\notag \\
    &\qquad \qquad \qquad  \quad \times \mathcal{N} (x ;u_{x,l_1},\sigma _{x,l_1}^{2}) \mathcal{N} (y;u_{y,l_2},\sigma _{y,l_2}^{2})\mathrm{d}x\mathrm{d}y \notag\\
    &=\sum\nolimits_{l_1=1}^{L_1}\sum\nolimits_{l_2=1}^{L_2}\phi _{x,l_1}^{}\phi _{y,l_2}^{}\iint_{\mathbb{R} ^{2\times 2}}J_{ij}(\mathbf{x};\mathbf{r})\mathcal{N} (x;u_{x,l_1},\sigma _{x,l_1}^{2}) \notag \\
    &\qquad \qquad \qquad \qquad \qquad \qquad \qquad \times \mathcal{N} (y;u_{y,l_2},\sigma _{y,l_2}^{2})\mathrm{d}x\mathrm{d}y \notag \\
    &=\frac{1}{2\pi}\sum\nolimits_{l_1=1}^{L_1}\sum\nolimits_{l_2=1}^{L_2}\phi _{x,l_1}^{}\phi _{y,l_2}^{}\iint_{\mathbb{R} ^{2\times 2}}J_{ij}(\mathbf{x};\mathbf{r})\frac{1}{\sigma _{x,l_1}^{} \sigma _{y,l_2}^{}}\notag \\
    &\qquad \qquad \qquad \times e^{-\frac{\left( x-u_{x,l_1} \right) ^2}{2\sigma _{x,l_1}^{2}}}e^{-\frac{\left( y-u_{y,l_2} \right) ^2}{2\sigma _{y,l_2}^{2}}}\mathrm{d}x\mathrm{d}y. \tag{A-1} \label{eq:A-1}
    \end{align} 
    Then, relying on the variable substitution $x=\sqrt{2}\sigma _{x,l_1}^{}\tilde{x}+u_{x,l_1}$ and $y=\sqrt{2}\sigma _{y,l_2}^{}\tilde{y}+u_{y,l_2}$, \eqref{eq:A-1} can be further written as 
    \begin{align}
        \eqref{eq:A-1}&=\frac{1}{\pi}\sum\nolimits_{l_1=1}^{L_1}{\sum\nolimits_{l_2=1}^{L_2}{\phi _{x,l_1}^{}\phi _{y,l_2}^{}}} \notag 
        \\ &\qquad \qquad\times \iint_{\mathbb{R} ^{2\times 2}}{J_{ij}(\mathbf{x};\mathbf{r})e^{-\tilde{x}^2}e^{-\tilde{y}^2}\mathrm{d}\tilde{x}\mathrm{d}\tilde{y}}, \notag
    \end{align}
    which aligns with the form that GHQ can address.
    Thus, we have the following derivations:
    \begin{align}
        &\eqref{eq:A-1} \approx \frac{1}{\pi}\sum\nolimits_{l_1=1}^{L_1}{\sum\nolimits_{l_2=1}^{L_2}{}}\phi _{x,l_1}^{}\phi _{y,l_2}^{} \times \notag \\
        &\sum_{i_1=1}^{I_1}{\sum_{i_2=1}^{I_2}{w_{i_1}w_{i_2}J_{ij}(\mathbf{x};\sqrt{2}\sigma _{x,l_1}^{}\tilde{x}_{i_1}+u_{x,l_1},\sqrt{2}\sigma _{y,l_2}^{}\tilde{y}_{i_2}+u_{y,l_2})}} \notag
        \\
        &=\frac{1}{\pi}\sum\nolimits_{l_1=1}^{L_1}{\sum\nolimits_{l_2=1}^{L_2}{\sum\nolimits_{i_1=1}^{I_1}{\sum\nolimits_{i_2=1}^{I_2}{\beta _{l_1,l_2,i_1,i_2}\notag }}}}\\
        &\times J_{ij}(\mathbf{x};\sqrt{2}\sigma _{x,l_1}^{}\tilde{x}_{i_1}+u_{x,l_1},\sqrt{2}\sigma _{y,l_2}^{}\tilde{y}_{i_2}+u_{y,l_2}), \tag{A-3}
    \end{align}
    where $\beta _{l_1,l_2,i_1,i_2}\triangleq \phi _{x,l_1}^{}\phi _{y,l_2}^{}w_{i_1}w_{i_2}$ denotes the weight of the GHQ.
    Here, this proof is completed.

\section{Derivations of the Optimal Power Allocation} \label{appendix_B}
According to \eqref{eq:bfim} and \eqref{eq:bcrb}, the closed-form expression for BCRB can be written as
\begin{align}
    \Gamma_{\rm sen} = I_{1,k} / I_{2,k}, \tag{B-1} \label{eq:B-1}
\end{align}
where 
\begin{align}
    &I_{1,k}=\alpha P\left( J_{yy}\left( \mathbf{x}_k \right) +J_{xx}\left( \mathbf{x}_k \right) \right) +\left( F_{yy,k}+F_{xx,k} \right) \notag \\
    &I_{2,k}=\alpha ^2P^2\left( J_{xx}\left( \mathbf{x}_k \right) J_{yy}\left( \mathbf{x}_k \right) -J_{xy}\left( \mathbf{x} \right) J_{yx}\left( \mathbf{x}_k \right) \right)\notag \\ 
    &+ \alpha P\left( J_{xx}\left( \mathbf{x}_k \right) F_{yy,k}+F_{xx,k}J_{yy}\left( \mathbf{x}_k \right) \right) +F_{xx,k}F_{yy,k}. \notag 
\end{align}
As $I_{2,k}>0$, \eqref{eq:B-1} can be written as 
\begin{align}
    P_k={-A_{2,k}+\sqrt{A_{2,k}^{2}-4A_{1,k}A_{3,k}}}/{(2A_{1,k})}, \tag{B-2} \label{eq:B-2}
\end{align}
where the terms are presented below:
\begin{align}
    &A_{1,k}\triangleq \alpha ^2\Gamma _{\mathrm{sen}}\left( J_{xx}\left( \mathbf{x}_k \right) J_{yy}\left( \mathbf{x}_k \right) -J_{xy}\left( \mathbf{x}_k \right) J_{yx}\left( \mathbf{x}_k \right) \right), \notag \\
    &A_{2,k}\triangleq \alpha \Gamma _{\mathrm{sen}}\left( J_{xx}\left( \mathbf{x}_k \right) F_{yy,k}+F_{xx,k}J_{yy}\left( \mathbf{x}_k \right) \right) \notag \\
    &\qquad -\alpha \left( J_{yy}\left( \mathbf{x}_k \right) +J_{xx}\left( \mathbf{x}_k \right) \right), \notag\\
    &A_{3,k}\triangleq \Gamma _{\mathrm{sen}}F_{xx,k}F_{yy,k}-\left( F_{yy,k}+F_{xx,k} \right). \notag  
\end{align}
In this case, the optimal power allocation policy can be obtained by keeping the positive root of \eqref{eq:B-2}.

\bibliographystyle{ieeetr}
\bibliography{mybib}

\begin{thebibliography}{10}

\bibitem{lu2014mimo}
L.~Lu, G.~Y. Li, A.~L. Swindlehurst, A.~Ashikhmin, and R.~Zhang, ``An overview
  of massive {MIMO}: Benefits and challenges,'' {\em IEEE J. Sel. Topics Signal
  Process.}, vol.~8, no.~5, pp.~742--758, Oct. 2014.

\bibitem{ouyang2025capacity}
C.~Ouyang, Z.~Wang, Y.~Liu, and Z.~Ding, ``Capacity characterization of
  pinching-antenna systems,'' {\em arXiv preprint arXiv:2506.14298}, 2025.

\bibitem{castellanos2025embracing}
M.~R. Castellanos, S.~Yang, C.-B. Chae, and R.~W.~H. Jr, ``Embracing
  reconfigurable antennas in the tri-hybrid {MIMO} architecture for {6G} and
  beyond,'' {\em arXiv preprint arXiv: 2501.16610}, 2025.

\bibitem{huang2019reconfigurable}
C.~Huang, A.~Zappone, G.~C. Alexandropoulos, M.~Debbah, and C.~Yuen,
  ``Reconfigurable intelligent surfaces for energy efficiency in wireless
  communication,'' {\em IEEE Trans. Wireless Commun.}, vol.~18, no.~8,
  pp.~4157--4170, Aug. 2019.

\bibitem{zhu2025tutorial}
L.~Zhu, W.~Ma, W.~Mei, {\em et~al.}, ``A tutorial on movable antennas for
  wireless networks,'' {\em IEEE Commun. Surv. Tut.}, 2025, early access,
  doi:10.1109/COMST.2025.3546373.

\bibitem{new2025tutorial}
W.~K. New, K.-K. Wong, H.~Xu, {\em et~al.}, ``A tutorial on fluid antenna
  system for {6G} networks: Encompassing communication theory, optimization
  methods and hardware designs,'' {\em IEEE Commun. Surv. Tut.}, vol.~27,
  no.~4, pp.~2325--2377, Aug. 2025.

\bibitem{new2024fluid}
W.~K. New, K.-K. Wong, H.~Xu, K.-F. Tong, and C.-B. Chae, ``Fluid antenna
  system: New insights on outage probability and diversity gain,'' {\em IEEE
  Trans. Wireless Commun.}, vol.~23, no.~1, pp.~128--140, Jan. 2024.

\bibitem{wong2022fluid}
K.-K. Wong and K.-F. Tong, ``Fluid antenna multiple access,'' {\em IEEE Trans.
  Wireless Commun.}, vol.~21, no.~7, pp.~4801--4815, Jul. 2022.

\bibitem{liu2022survey}
A.~Liu, Z.~Huang, M.~Li, Y.~Wan, W.~Li, T.~X. Han, C.~Liu, R.~Du, D.~K.~P. Tan,
  J.~Lu, Y.~Shen, F.~Colone, and K.~Chetty, ``A survey on fundamental limits of
  integrated sensing and communication,'' {\em IEEE Commun. Surv. Tut.},
  vol.~24, no.~2, pp.~994--1034, 2nd Quart. 2022.

\bibitem{ma2024movable}
W.~Ma, L.~Zhu, and R.~Zhang, ``Movable antenna enhanced wireless sensing via
  antenna position optimization,'' {\em IEEE Trans. Wireless Commun.}, vol.~23,
  no.~11, pp.~16575--16589, Nov. 2024.

\bibitem{liu2025pinching}
Y.~Liu {\em et~al.}, ``Pinching-antenna systems: Architecture designs,
  opportunities, and outlook,'' {\em IEEE Commun. Mag.}, pp.~1--7, 2025, early
  access, doi: 10.1109/MCOM.001.2500037.

\bibitem{ouyang2025array}
C.~Ouyang, Z.~Wang, Y.~Liu, and Z.~Ding, ``Array gain for pinching-antenna
  systems ({PASS}),'' {\em IEEE Commun. Lett.}, vol.~29, no.~6, pp.~1471--1475,
  Jun. 2025.

\bibitem{fukuda2022pinching}
A.~Fukuda, H.~Yamamoto, H.~Okazaki, Y.~Suzuki, and K.~Kawai, ``Pinching
  antenna: Using a dielectric waveguide as an antenna,'' {\em NTT DOCOMO Tech.
  J.}, vol.~23, pp.~5--12, Jan. 2022.

\bibitem{ding2025flexible}
Z.~Ding, R.~Schober, and H.~Vincent~Poor, ``Flexible-antenna systems: A
  pinching-antenna perspective,'' {\em IEEE Trans. Commun.}, 2025, early
  access, doi:10.1109/TCOMM.2025.3555866.

\bibitem{xu2025pinching}
Y.~Xu, Z.~Ding, R.~Schober, and T.-H. Chang, ``Pinching-antenna systems with
  in-waveguide attenuation: Performance analysis and algorithm design,'' {\em
  arXiv preprint arXiv:2506.23966}, 2025.

\bibitem{wang2025modeling}
Z.~Wang, C.~Ouyang, X.~Mu, Y.~Liu, and Z.~Ding, ``Modeling and beamforming
  optimization for pinching-antenna systems,'' {\em arXiv preprint
  arXiv:2502.05917}, 2025.

\bibitem{rusek2013scaling}
F.~Rusek, D.~Persson, B.~K. Lau, E.~G. Larsson, T.~L. Marzetta, O.~Edfors, and
  F.~Tufvesson, ``Scaling up {MIMO}: Opportunities and challenges with very
  large arrays,'' {\em IEEE Signal Process. Mag.}, vol.~30, no.~1, pp.~40--60,
  Jan. 2013.

\bibitem{jiang2025nearfield}
H.~Jiang, Z.~Wang, Y.~Liu, H.~Shin, A.~Nallanathan, and Y.~Liu, ``Near-field
  sensing enabled predictive beamforming: Fundamentals, framework, and
  opportunities,'' {\em arXiv preprint arXiv:2506.09225}, 2025.

\bibitem{qin2025joint}
Y.~Qin, Y.~Fu, and H.~Zhang, ``Joint antenna position and transmit power
  optimization for pinching antenna-assisted {ISAC} systems,'' {\em arXiv
  preprint arXiv:2503.12872}, 2025.

\bibitem{zhang2025integrated}
Z.~Zhang {\em et~al.}, ``Integrated sensing and communications for
  pinching-antenna systems ({PASS}),'' {\em arXiv preprint arXiv:2504.07709},
  2025.

\bibitem{kay1993estimation}
S.~M. Kay, {\em Fundamentals of Statistical Signal Processing: Estimation
  Theory}.
\newblock Englewood Cliffs, NJ, USA: Prentice-Hall, 1993.

\bibitem{liu2022cramer}
F.~Liu, Y.-F. Liu, A.~Li, C.~Masouros, and Y.~C. Eldar, ``Cram{é}r-{R}ao bound
  optimization for joint radar-communication beamforming,'' {\em IEEE Trans.
  Signal Process.}, vol.~70, pp.~240--253, 2022.

\bibitem{ding2025pinchingisac}
Z.~Ding, ``Pinching-antenna assisted {ISAC}: A {CRLB} perspective,'' {\em arXiv
  preprint arXiv:2504.05792}, 2025.

\bibitem{wang2025wireless}
Z.~Wang, C.~Ouyang, Y.~Liu, and A.~Nallanathan, ``Wireless sensing via
  pinching-antenna systems,'' {\em arXiv preprint arXiv: 2505.15430}, 2025.

\bibitem{bozanis2025cramer}
D.~Bozanis, V.~K. Papanikolaou, S.~A. Tegos, and G.~K. Karagiannidis,
  ``{C}ram\'er-{R}ao bounds for integrated sensing and communications in
  pinching-antenna systems,'' {\em arXiv preprint arXiv:2505.01333}, 2025.

\bibitem{li2025pinching}
H.~Li, R.~Zhong, J.~Lei, and Y.~Liu, ``Pinching antenna system for integrated
  sensing and communications,'' 2025.

\bibitem{khalili2025pinching}
A.~Khalili, B.~Kaziu, V.~K. Papanikolaou, and R.~Schober, ``Pinching
  antenna-enabled {ISAC} systems: Exploiting look-angle dependence of rcs for
  target diversity,'' {\em arXiv preprint arXiv:2505.01777}, 2025.

\bibitem{ouyang2025rate}
C.~Ouyang, Z.~Wang, Y.~Liu, and Z.~Ding, ``Rate region of {ISAC} for
  pinching-antenna systems,'' {\em arXiv preprint arXiv:2505.10179}, 2025.

\bibitem{xu2024mimo}
C.~Xu and S.~Zhang, ``{MIMO} integrated sensing and communication exploiting
  prior information,'' {\em IEEE J. Sel. Areas Commun.}, vol.~42, no.~9,
  pp.~2306--2321, Sept. 2024.

\bibitem{hou2024optimal}
K.~Hou and S.~Zhang, ``Optimal beamforming for secure integrated sensing and
  communication exploiting target location distribution,'' {\em IEEE J. Sel.
  Areas Commun.}, vol.~42, no.~11, pp.~3125--3139, Nov. 2024.

\bibitem{dauwels2005computing}
J.~Dauwels, ``Computing {Bayesian} {C}ram{\'e}r-{R}ao bounds,'' in {\em Proc.
  IEEE Int. Symp. Inf. Theory (ISIT)}, pp.~425--429, Sept. 2005.

\bibitem{ding2025pinching}
Z.~Ding, ``Pinching-antenna assisted {ISAC}: A {CRLB} perspective,'' {\em arXiv
  preprint arXiv: 2504.05792}, 2025.

\bibitem{jiang2024cramer}
H.~Jiang, Z.~Wang, Y.~Liu, and A.~Nallanathan, ``{Cram\'er-Rao} bound
  optimization for near-field sensing with continuous-aperture arrays,'' {\em
  arXiv preprint arXiv: 2412.15007}, 2024.

\bibitem{li2008range}
J.~Li, L.~Xu, P.~Stoica, K.~W. Forsythe, and D.~W. Bliss, ``Range compression
  and waveform optimization for mimo radar: A cram{é}r–rao bound based
  study,'' {\em IEEE Trans. Signal Process.}, vol.~56, no.~1, pp.~218--232,
  Jan. 2008.

\bibitem{kay1998fundamentals}
S.~M. Kay, {\em Fundamentals of Statistical Signal Processing: Estimation
  Theory}.
\newblock Englewood Cliffs, NJ, USA: Prentice Hall, 1998.

\bibitem{van2004detection}
H.~L. Van~Trees, {\em Detection, Estimation, and Modulation Theory, Part {I}:
  Detection, Estimation, and Linear Modulation Theory}.
\newblock New York, NY, USA: John Wiley \& Sons, 2004.

\bibitem{scope2025bayesian}
E.~Scope~Crafts, X.~Zhang, and B.~Zhao, ``Bayesian {C}ram{é}r-{R}ao bound
  estimation with score-based models,'' {\em IEEE Trans. Inf. Theory}, vol.~71,
  no.~3, pp.~2007--2027, Mar. 2025.

\bibitem{shen2010fundamental}
Y.~Shen and M.~Z. Win, ``Fundamental limits of wideband localization— part
  {I}: A general framework,'' {\em IEEE Trans. Inf. Theory}, vol.~56, no.~10,
  pp.~4956--4980, Oct. 2010.

\bibitem{nguyen2023approximation}
T.~Nguyen, F.~Chamroukhi, H.~D. Nguyen, and G.~J. McLachlan, ``Approximation of
  probability density functions via location-scale finite mixtures in lebesgue
  spaces,'' {\em Commun. Statistics-Theory and Methods}, vol.~52, no.~14,
  pp.~5048--5059, 2023.

\bibitem{wang2025beamfocusing}
Z.~Wang, X.~Mu, and Y.~Liu, ``Beamfocusing optimization for near-field wideband
  multi-user communications,'' {\em IEEE Trans. Commun.}, vol.~73, no.~1,
  pp.~555--572, Jan. 2025.

\end{thebibliography}

\end{document}